\documentclass[preprint2,numberedappendix]{emulateapj-rtx4}
\usepackage{graphicx,natbib}
\usepackage{bm,url,color}
\usepackage{amsmath}
\usepackage{amssymb}
\usepackage{graphics}
\usepackage{euscript}
\graphicspath{{./figures/}{./png/}}
\graphicspath{{./fig2/}{./fig3/}}

\newcommand{\EQ}{\begin{equation}}
\newcommand{\EN}{\end{equation}}
\newcommand{\EQA}{\begin{eqnarray}}
\newcommand{\ENA}{\end{eqnarray}}

\newcommand{\Eq}[1]{Equation~(\ref{#1})}

\newcommand{\Sec}[1]{Section~\ref{#1}}

\newcommand{\Fig}[1]{Figure~\ref{#1}}

\newcommand{\Figs}[2]{Figures~\ref{#1} and \ref{#2}}

\newcommand{\Figp}[2]{Figure~\ref{#1}({#2})}
\newcommand{\Figsp}[3]{Figures~\ref{#1}({#2}) and ({#3})}

\newcommand{\bra}[1]{\langle #1\rangle}

{}
{}
{}

{}
{}
{}
{}
{}
{}
{}
{}
{}
{}
{}
{}
{}
{}
{}
{}
{}
{}
{}
{}
{}
{}

{}
{}
{}

{}

{}
{}

\newcommand{\dd}{{\rm d} {}}

\def\degr{\hbox{$^\circ$}}

\def\kurt{\mbox{\rm kurt\,}}

\def\Brms{B_{\rm rms}}

\newcommand{\G}{\,{\rm G}}

\newcommand{\nHz}{\,{\rm nHz}}

\newcommand{\mHz}{\,{\rm mHz}}

\newcommand{\kG}{\,{\rm kG}}

\newcommand{\kmeter}{\,{\rm km}}
\newcommand{\h}{\,{\rm h}}
\newcommand{\s}{\,{\rm s}}

\newcommand{\m}{\,{\rm m}}

\newcommand{\Mm}{\,{\rm Mm}}

\newcommand{\yapj}[3]{ #1, {ApJ,} {#2}, #3}

\newcommand{\yapjl}[3]{ #1, {ApJ,} {#2}, #3}

\newcommand{\yarep}[3]{ #1, {ARep,} {#2}, #3}
\newcommand{\yana}[3]{ #1, {A\&A,} {#2}, #3}

\newcommand{\yanar}[3]{ #1, {A\&A Rev.,} {#2}, #3}

\newcommand{\ymn}[3]{ #1, {MNRAS,} {#2}, #3}
\newcommand{\ynat}[3]{ #1, {Nature,} {#2}, #3}
\newcommand{\ysci}[3]{ #1, {Science,} {#2}, #3}
\newcommand{\ysph}[3]{ #1, {Solar Phys.,} {#2}, #3}

\newcommand{\yjour}[4]{ #1, {#2}, {#3}, #4}

\newcommand{\beq}{\begin{equation}}
\newcommand{\eeq}{\end{equation}}

\begin{document}

\title{High-wavenumber solar \MakeLowercase{\mbox{$f$}}-mode
strengthening prior to active region formation}
\author{Nishant K. Singh$^1$, Harsha Raichur$^1$, \&
Axel Brandenburg$^{1,2,3,4}$}
\affil{$^1$Nordita, KTH Royal Institute of Technology and Stockholm University, Roslagstullsbacken 23, SE-10691 Stockholm, Sweden\\
$^2$JILA and Department of Astrophysical and Planetary Sciences, University of Colorado, Boulder, CO 80303, USA\\
$^3$Department of Astronomy, AlbaNova University Center, Stockholm University, SE-10691 Stockholm, Sweden\\
$^4$Laboratory for Atmospheric and Space Physics, University of Colorado, Boulder, CO 80303, USA
}

\submitted{\today,~ $ $Revision: 1.318 $ $}

\begin{abstract}
We report a systematic strengthening of the local solar surface or
fundamental $f$-mode $1$--$2$ days prior to the emergence of an active
region (AR) in the same (corotating) location.
Except for a possibly related increase in the kurtosis of the magnetic
field, no indication can be seen in the magnetograms at that time.
Our study is motivated by earlier numerical findings of
\cite{SBR14} which showed that, in the presence of a nonuniform magnetic field
that is concentrated a few scale heights below the surface, the $f$-mode
fans out in the diagnostic $k\omega$ diagram at high wavenumbers.
Here we explore this possibility using data from the Helioseismic and
Magnetic Imager on board the {\em Solar Dynamics Observatory}
and show for six isolated ARs, 11130, 11158, 11242, 11105, 11072, and 11768,
that at large latitudinal
wavenumbers (corresponding to horizontal scales of around $3000\kmeter$),
the $f$-mode displays strengthening
about two days prior to AR formation and thus provides a new precursor for
AR formation.
Furthermore, we study two ARs, 12051 and 11678, apart from a magnetically
quiet patch lying next to AR~12529, to demonstrate the challenges
in extracting such a precursor signal when a newly forming AR emerges
in a patch that lies in close proximity of one or several already
existing ARs which are expected to pollute neighboring patches.
We then discuss plausible procedures for extracting precursor
signals from regions with crowded environments.
The idea that the $f$-mode is perturbed days before any visible
magnetic activity occurs at the surface can be important in
constraining dynamo models aimed at understanding
the global magnetic activity of the Sun.
\end{abstract}

\email{nishant@nordita.org}
\section{Introduction}

Recent work has demonstrated the potential usefulness of employing the
surface or fundamental $f$-mode in local helioseismology for detecting
subsurface solar magnetism \citep{HBBG08,DABCG11,FBCB12,FCB13}.
While turbulence generally tends to lower the $f$-mode frequency
\citep{FSTT92,MR93b,DKM98} relative to its theoretical value given by
$\omega_f=\sqrt{gk}$, where $g$ is the gravitational acceleration and $k$
is the horizontal wavenumber, horizontal magnetic fields can increase the
frequency \citep{MR93a}, while vertical or inclined fields lead to a nonuniform
behavior, depending on the value of the horizontal wavenumber \citep{SBCR15}.
More importantly, however, horizontal {\em variability} of the subsurface
magnetic field leads to a fanning of the $f$-mode, where changes in the
integrated mode amplitude and position give clues about the depth of
such a field \citep{SBR14}.
While these investigations demonstrated a number of previously unknown
effects of the $f$-mode, they were restricted to idealizing conditions
of an isothermal layer.
In this Letter, we use observations with the Helioseismic and
Magnetic Imager (HMI) on board the {\em Solar Dynamics Observatory}
({\em SDO}) to search for possible similarities between observations
and simulations.

We focus on the possibility of using changes in the $f$-mode
to predict the emergence of active regions (ARs) days before they can be seen
in magnetograms.
Owing to the very nature of the $f$-mode
being confined to the proximity of the surface,
our approach is most sensitive to magnetic fields at
shallow depths of just a few megameters (Mm), and ceases to
be sensitive when the AR begins to become fully developed.
Earlier attempts of predicting the emergence of ARs employed
time-distance seismology using $p$-modes and have suggested the occurrence
of perturbations at larger depths of $40$--$75\Mm$ \citep{Ilo11,Kho13}.
On the other hand, the rising flux tube scenario suggests a retrograde flow
at a depth of $30\Mm$ \citep{BBF10}, which has not been observed
\citep{Birch16}.
Also morphological studies in the case of AR~11313 have suggested
incompatibilities with the rising flux tube model \citep{Get16}.
By contrast, in the distributed dynamo scenario \citep{B05}, magnetic
flux concentrations form spontaneously near
the surface \citep{BKKMR11,BKR13}, which might explain
the aforementioned field concentrations at shallow depths.
Spontaneous surface flux concentrations have also been seen in the deep
hydromagnetic convection simulations of \cite{SN12}, where an unstructured
magnetic field is allowed to enter the bottom of their computational domain.
Such near-surface magnetic concentrations are expected to affect the
$f$-mode as its eigenfunction peaks only a few $\Mm$
below the solar surface \citep[cf.][]{Sch99}.
It is possible that these perturbations could manifest themselves
through detectable signatures.

Readers familiar with the conventional picture of buoyant flux tube emergence
\citep[as reviewed by, e.g.,][]{Cha10} might be concerned about depths
as shallow as just a few Mm, because buoyant tubes of several kilogauss
would reach the surface within an hour \citep[$\sim$~3 hours from the
depth of 7.5 Mm in the simulations of][]{CRTS10},
but this picture ignores the formation process and implants flux tubes
as alien objects within the turbulent convection zone.
By contrast, ARs and sunspots might instead be generated by the subsurface
turbulence in ways similar to what has so far only been seen in
idealized simulations \citep{BKR13,WLBKR13,MBKR14}.
The point here is not to defend this idea, but to raise awareness
of alternative viewpoints that would facilitate the understanding of
the results presented below in the present work.

Once the AR has been detected in magnetograms and becomes
fully developed, the $f$-mode amplitude begins to be suppressed.
This might be explained by the fact that 
the interaction of both $f$- and $p$-modes with
ARs or sunspots leads to mode conversion, resulting in
the absorption of mode power \citep{TCN82,CBZ94,CB97}.
This would explain the observed reduction of the mode amplitude
\emph{after} the analyzed AR has been formed.
However, what was not discussed earlier is that the mode amplitude
from the same region can undergo
a {\em transient} growth phase prior to the actual flux emergence.
This results in a nonmonotonic temporal variation in the
normalized mode power which first rises, reaches a maximum value
a few days before there is any sign of flux emergence, and then
decreases as the strength of the magnetic field in that region increases.
Although a proper explanation of this is not yet available, one might
speculate that this could also be due to some kind of scattering,
whereby $p$-modes would scatter off the magnetic flux concentrations
and leak into enhanced $f$-mode power.

\section{Data analysis}

We use line-of-sight (LoS) Dopplergrams and magnetograms
from observations with HMI, mostly in the cylindrical equal-area
projection mappings that are publicly available on the Joint Science
Operations Center at Stanford\footnote{\url{http://jsoc.stanford.edu/}}.
Our analysis is based on 45 seconds cadence data with a projection
scale of $0.03\degr$ per pixel, where the data represent the
LoS Doppler velocity $v(x,y,t)$ as a function of horizontal position
$(x,y)$ and time $t$.
For each of the regions of interest, we consider a patch of
$512^2$ pixels covering an area of about $(180\Mm)^2$
$\approx(15\degr)^2$ on the solar surface. We track these patches
for several days using a frame of reference corotating with the mean
(Carrington) rotation rate $\Omega_0$ with $\Omega_0/2\pi=424\nHz$.
To capture transient signatures, we use data cubes
$v(x,y,t)$ of only 8 hours duration
for the entire tracking period of our target region.
To reduce the noise level
arising from solar convection \citep{Zha15} and effects from latitudinal
differential rotation (J.\ Zhao, private communication),
we use a running difference to the
original images before storing $v(x,y,t)$.

\begin{figure}
\begin{center}
\includegraphics[width=\columnwidth]{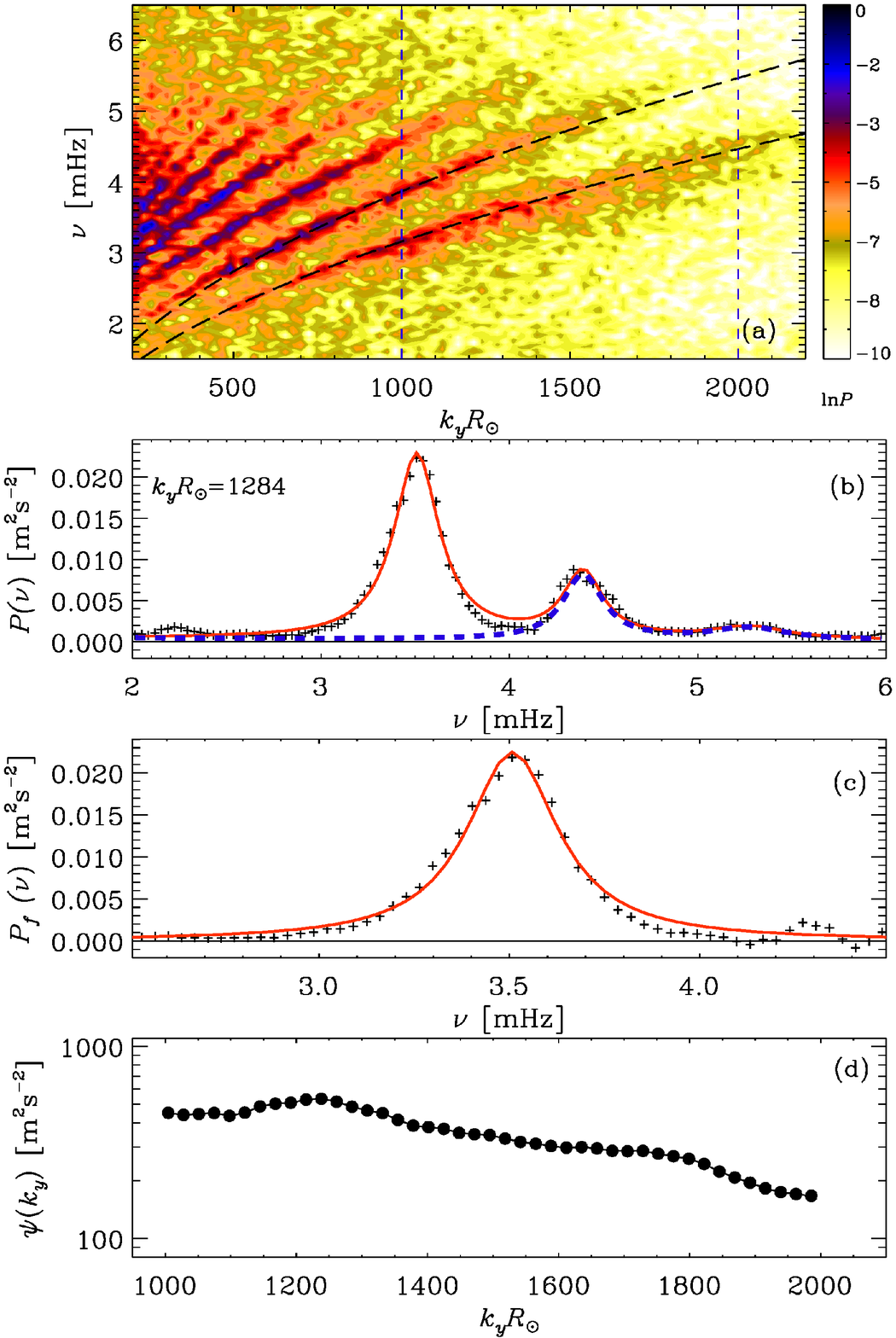}
\end{center}
\caption[]{
(a) Typical $k\omega$ diagram where the lowest ridge is the $f$-mode,
here for the quiet sun during 2010 May 14;
(b) example of a vertical cut at a specified value of
$k_y R_\odot$ (plus symbols) together with the model fit (solid, red curve)
and $P_{\rm cp}$ (dashed, blue line);
(c) $f$-mode ridge ($P_f$, plus symbols) and the corresponding fit
(solid, red curve); 
(d) $\psi(k_y)$ for the full range enclosed within
the vertical dashed lines in (a).}
\label{QS2010_kyo_psi}
\end{figure}

We divide our five or six day stretches into 15 or 18 intervals of 8 hrs,
each resulting in a data cube of $512^2\times640$ points of
$v(x,y,t)$ that is Fourier transformed to give $\hat{v}(k_x,k_y,\omega)$,
which too has the dimension $\m\s^{-1}$ in our normalization.
We then construct power spectra from $P=|\hat{v}|^2$ and select
$k_x=0$ in the subsequent analysis.
Thus, we ignore longitudinal variations that could be affected by
the cylindrical equal-area projection, as the latitudinal
directions are expected to be the least sensitive to artifacts resulting
from projection and also differential rotation.
Also, our target regions were chosen such that the patches were always far
from the limb during the entire tracking period.
The thus obtained power spectra $P(k_x=0,k_y,\omega)$
are then used to construct
a diagnostic $k\omega$ diagram in the $k_y$-$\omega$ plane;
see \Figp{QS2010_kyo_psi}{a} which displays the
$f$- and $p$-ridges where the horizontal wavenumber is $k=k_y$.

We now take a cut parallel to the frequency axis at a fixed $k_y R_\odot$
to get the line profiles of the $f$- and lowest two $p$-ridges.
We then apply boxcar smoothing along the frequency axis with
a box width of $0.24\mHz$.
To determine the strength of the $f$-mode,
we remove first the continuum and the lowest two $p$-ridges, which are
represented by a superposition of parabolic and Lorentzian fits,
respectively and denoted by $P_{\rm cp}=|\hat{v}|^2_{\rm cp}$,
where the subscript cp stands for the sum of continuum and $p$-modes;
see \Figsp{QS2010_kyo_psi}{b}{c}.
In most cases we repeat the same procedure at all wavenumbers in
the range $k_y R_\odot \in [1200, 2000]$,
and determine the $f$-mode power as
$P_f(k_y,\omega)=|\hat{v}_f|^2=P-P_{\rm cp}$.
We may define the integrated $f$-mode amplitude assuming circularly
symmetric rings in the $k_x$-$k_y$ plane as
\begin{equation}
\bra{v^2}_f=2AT\int_0^\infty\int_0^\infty k P_f(k,\omega)\,
{\dd k\over2\pi}\, {\dd\omega\over2\pi},
\label{v2f}
\end{equation}
where $k^2=k_x^2+k_y^2$, and write $\bra{v^2}_f$ as
\begin{equation}
\bra{v^2}_f=L\sum_{k}\; k P_{f,k}
\; \mbox{with}\;
P_{f,k} = 2 \sum_{\omega} P_f(k,\omega),
\label{v2f2}
\end{equation}
where $A=L^2$ is the area of the chosen patch, $L$ is the side length
and $T$ is the tracking time of the data cube.
Thus, we can determine the energy of the $f$-mode, $E_f$, characterizing
its strength, as:
\begin{equation}
E_f(t)\equiv{1\over2}\bra{v^2}_f(t)={1\over2}\left(L\over R_\odot\right)
\sum_{k}\; \psi(k)
\label{ef}
\end{equation}
with $\psi(k)=k R_\odot\, P_{f,k}$; see \Figp{QS2010_kyo_psi}{d}.
Note that we determine the above quantities by setting $k_x=0$
and choosing a high-wavenumber range, $k_y R_\odot \in [1200,2000]$,
unless otherwise specified. Although this choice of considering only
high wavenumbers in assessing the strength of the mode is
not a standard procedure, we nevertheless focus on this regime as
this ``precursor signal'' appears to be localized at such
large wavenumbers; see \Sec{results} below.
The time dependence of $E_f$ may now be determined by computing the
above quantities from the sequence of $8\h$ data cubes prepared for
all tracked regions of interests.

Even in the quiet phase during solar minimum,
$E_f$ shows a systematic dependence on
the angular distance $\alpha$ from the disk center,
given by
\begin{equation}
\cos\alpha=\cos\vartheta\cos\varphi\;;\quad
\varphi=\varphi_*-\varphi_0+\Omega_{\rm syn}t,
\label{cos}
\end{equation}
with $\vartheta$ and $\varphi$ being respectively the latitude and
longitude of the point of interest, $\varphi_*$ is the corresponding
Carrington longitude, $\varphi_0$ is the Carrington longitude of the
disk center at the time when we began tracking the target patch, and
$\Omega_{\rm syn}=2\pi/27.275\,{\rm days}$ is the mean
synodic Carrington rotation rate of the Sun (i.e., the apparent rotation
rate as viewed from the Earth).

As suggested by earlier work \citep{SBR14}, we focus on $E_f$
for fairly large $k_y$.
We track a particular position on the solar surface in time using the
average (Carrington) rotation rate.
Normalizing by the solar radius $R_\odot=700\Mm$ gives the spherical
harmonic degree $k_y R_\odot$.
For a fixed range of $k_y R_\odot$,
we compute the dependence of $E_f$ on $t$.
Empirically, the value of $E_f$ for the quiet sun (the position where
no AR emerges within the next
few days) shows a systematic variation that is approximately of the
form
\EQ
\zeta(\cos\alpha)=\cos\alpha\left[q+(1-q)\cos\alpha\right]\;
\mbox{with}\;q=0.5.
\EN
This function obeys $\zeta=1$ at $\alpha=0$ (disk center).
It is then useful to define
\EQ
\widetilde{E}_f\equiv E_f/\zeta,
\EN
which fluctuates moderately about
some average value in the quiet phase of the Sun.
However, several days {\em prior} to the emergence of an AR, our studies show
elevated values of $\widetilde{E}_f$ at that corotating patch where this
AR later emerges.

It would be interesting to see whether there are other indicators,
for example in the magnetic field itself, which could also give early
indications of AR formation.
Magnetic properties from regions of interest on the solar disk
might offer insight into the process of developing ARs. The
LoS magnetic field ($B$) varies randomly in space
and time, and has a narrow distribution with positive and negative
polarities nearly balancing themselves out when the localized patch
is magnetically quiet. Let us denote by $f_B$ the normalized
probability distribution function (PDF) of $B$ in a chosen patch
at any given time, such that
\EQ
\int_{-\infty}^\infty f_B \, \dd B =1.
\EN
The kurtosis, $\kurt B$, of the distribution $f_B$ is defined as,
\EQ
\kurt B=\frac{1}{\sigma_B^4}\int_{-\infty}^\infty
\left(B-\overline{B}\right)^4 f_B \, \dd B,
\label{kurt}
\EN
where the mean ($\overline{B}$) and the variance ($\sigma_B$) of $f_B$ are
\EQ
\overline{B}=\int_{-\infty}^\infty B f_B \, \dd B\;\;\mbox{and}\;\;
\sigma_B^2=\int_{-\infty}^\infty (B-\overline{B})^2 f_B \, \dd B,
\EN
respectively.
For a normal distribution, $\kurt B=3$, while excess kurtosis,
$\kurt B\gg3$, indicates a heavy-tailed distribution.
We monitor the temporal evolution of $\kurt B$ from the localized
patches that we track on the solar disk as the Sun rotates.

It is useful to
make a simultaneous comparison with the value for relatively
quiet patches under otherwise identical local conditions.
This may be realized as follows:
corresponding to each target region at $(\vartheta, \varphi)$,
we consider a (quiet) mirror
region at $(\vartheta^\dag, \varphi)$ in the opposite hemisphere
with the same dimensions, and track both these patches simultaneously,
where $\vartheta^\dag=-\vartheta$ for the
entire tracking period.
We refer to the $f$-mode energy from such a mirror region as $E_f^\dag$.
We find that, while the rms magnetic field $\Brms$ rises when the AR
emerges, the value in the mirror region, $\Brms^\dag$, remains close to
a constant background value.

\section{Sample selection}

We have selected a number of ARs, which may be broadly
classified under the following two categories:
\begin{itemize}
\item \emph{Isolated ARs:} In these examples, ideally
a single AR emerges in isolation, with the rest of
the Sun being nearly magnetically quiet.
As the seismic signals may well be nonlocal, we first need
to study isolated ARs to assess the $f$-mode perturbation due to
subsurface magnetic fields associated with newly developing ARs.
This would allow us to avoid contamination that might
be caused by the presence of already existing ARs in the
neighborhood of the patch where a new AR is going to form later.
There are not many instances since the launch of {\em SDO}, where
only a single AR appears on the entire solar disk, and therefore we
have included a few more cases wherein the other ARs are at least far
($>300\Mm$) from the AR in emergence.
The chosen examples in this class include ARs 11130, 11158, 11242,
11105, 11072 and 11768.
\item \emph{Crowded ARs:} It would be a serious limitation if the
proposed technique applies only to isolated ARs, and therefore we
have also studied the effects of a newly forming AR on a patch
that lies in close proximity to one or several already existing ARs.
To highlight the challenges one might face in extracting the signal
from such an AR, we have studied ARs 12051 and 11678.
\end{itemize}
Furthermore, in order to avoid systematic effects close to the limb,
we have restricted our sample to only those cases
which lie within $\pm60\degr$ in both latitude and longitude
of the disk center.
Yet another requirement limiting our sample size
is that the corresponding mirror patches in the opposite hemisphere
must be magnetically quiet for the entire tracking
period, thus offering an easy and simultaneous control.

We also studied four magnetically quiet patches at two different phases
of the solar magnetic activity cycle.
Two such patches, symmetrically located in the northern and the southern
hemispheres, were chosen when the Sun was just coming out of its minimum
during 2010 May. This offers another control when the Sun
did not show much magnetic activity for a few days.
We then chose a magnetically quiet patch lying next to AR~12529
during 2016 April, and also followed simultaneously its mirror
counterpart.

\begin{figure}
\begin{center}
\includegraphics[width=\columnwidth]{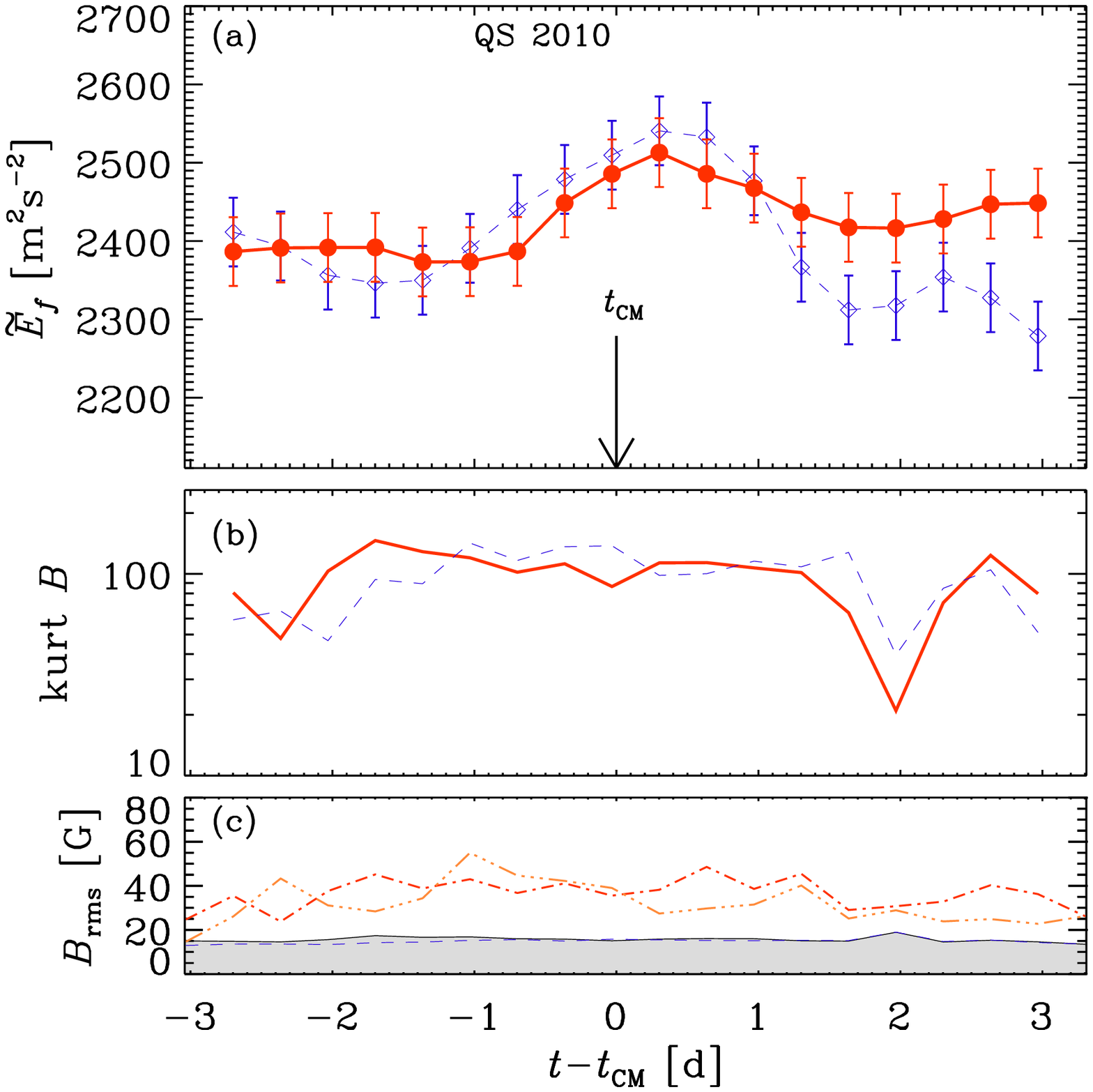}
\includegraphics[width=\columnwidth]{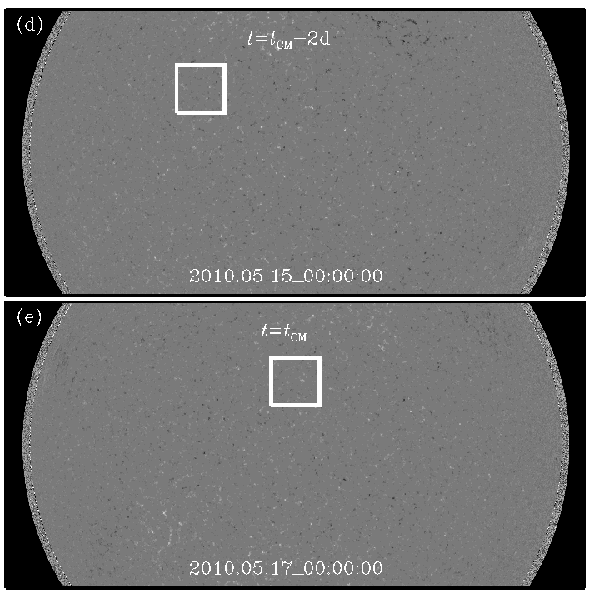}
\end{center}
\caption[]{
Time traces of $\widetilde{E}_f$ (solid red; $\vartheta=+20\degr$)
and $\widetilde{E}_f^\dag$ (dashed blue; $\vartheta=-20\degr$)
as a function of $t-t_{\rm CM}$ in panel (a),
evolutions of the kurtosis, $\kurt B$ (solid red) and
$\kurt B^\dag$ (dashed blue) in panel (b),
$\Brms$ (solid line with shaded area underneath) together with $\Brms^\dag$
(dashed blue line) in panel (c), as well as
magnetograms at $t=t_{\rm CM}-2{\rm d}$ (d) and $t=t_{\rm CM}$
(e) for the quiet sun during 2010 May 14--19.
The dash-dotted (red) and triple-dot-dashed (orange)
lines denote the time-traces of $0.08\,B_{\rm max}$ and $-0.08\,B_{\rm min}$,
respectively from the patch in the northern hemisphere.
}\label{t-trace_QS2010}
\end{figure}

\begin{figure}
\begin{center}
\includegraphics[width=\columnwidth]{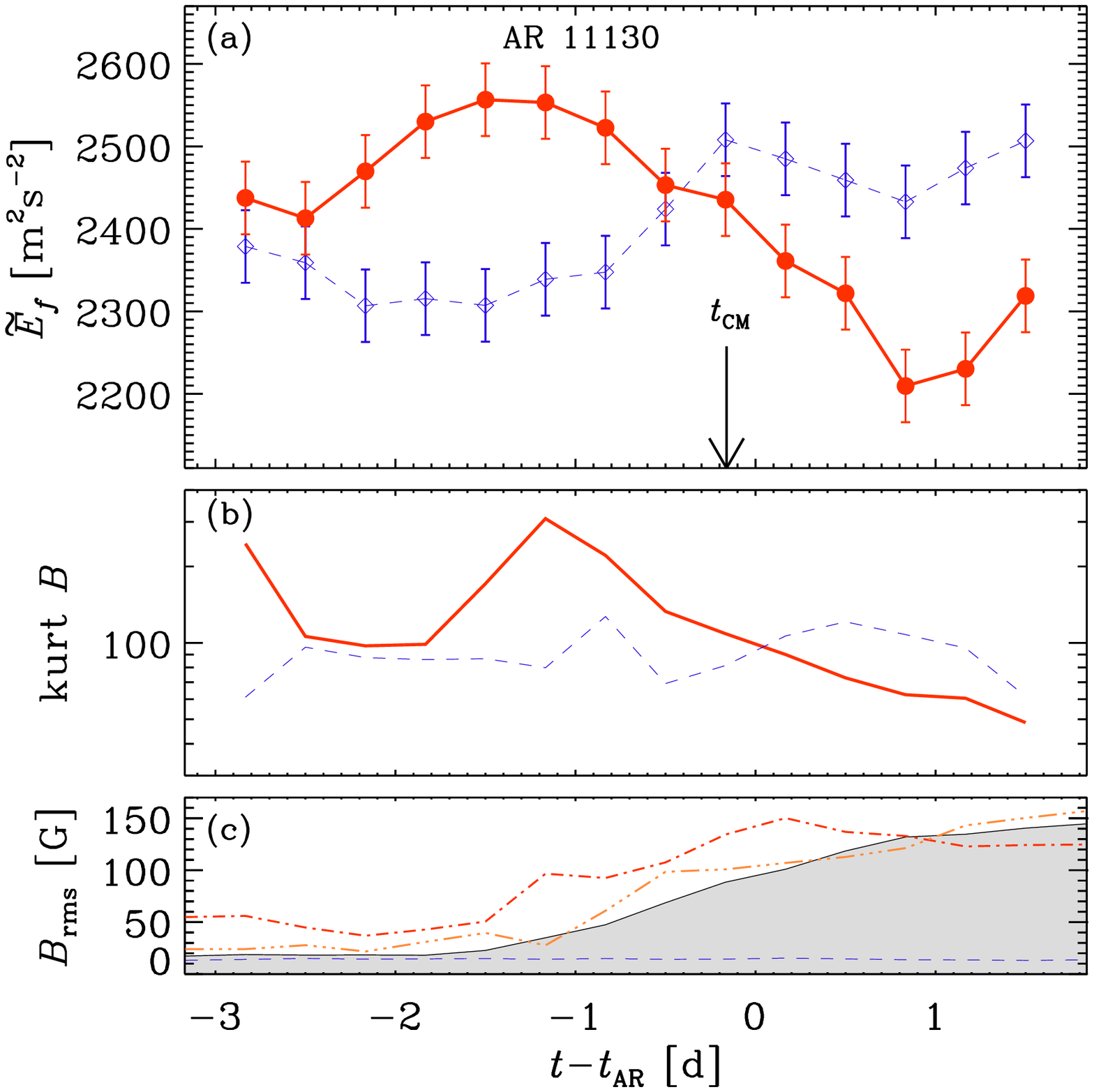}
\includegraphics[width=\columnwidth]{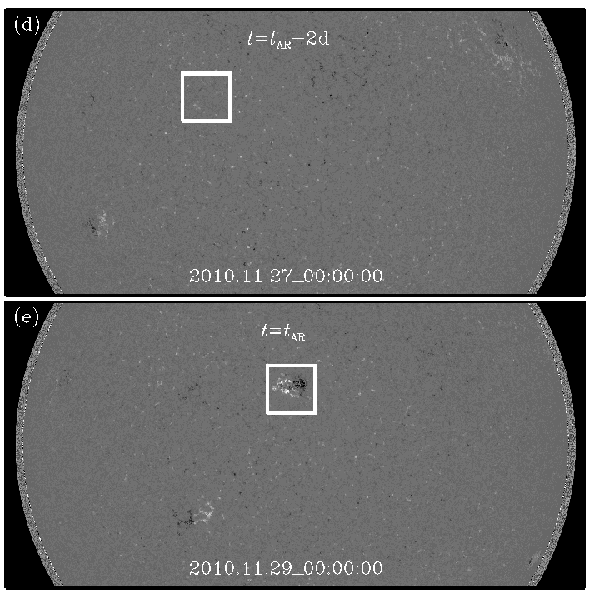}
\end{center}
\caption[]{
Time traces of $\widetilde{E}_f$ (solid red line) and $\widetilde{E}_f^\dag$
(dashed blue line) as a function of $t-t_{\rm AR}$, with $t_{\rm CM}$
marking the time of central meridian crossing in panel (a),
evolutions of the excess kurtosis, $\gamma_B$ (solid red) and
$\gamma_B^\dag$ (dashed blue) in panel (b),
$\Brms$ (solid line with shaded area underneath) together with $\Brms^\dag$
(dashed blue line) as a function of $t-t_{\rm AR}$ (c), as well as
magnetograms at $t=t_{\rm AR}-2{\rm d}$ (d) and $t=t_{\rm AR}$
(e) for AR~11130. The dash-dotted (red) and triple-dot-dashed (orange)
lines denote the time-traces of $0.08\,B_{\rm max}$ and $-0.08\,B_{\rm min}$,
respectively from the patch where the AR~11130 develops.
}\label{t-trace_11130}
\end{figure}

\begin{figure}
\begin{center}
\includegraphics[width=\columnwidth]{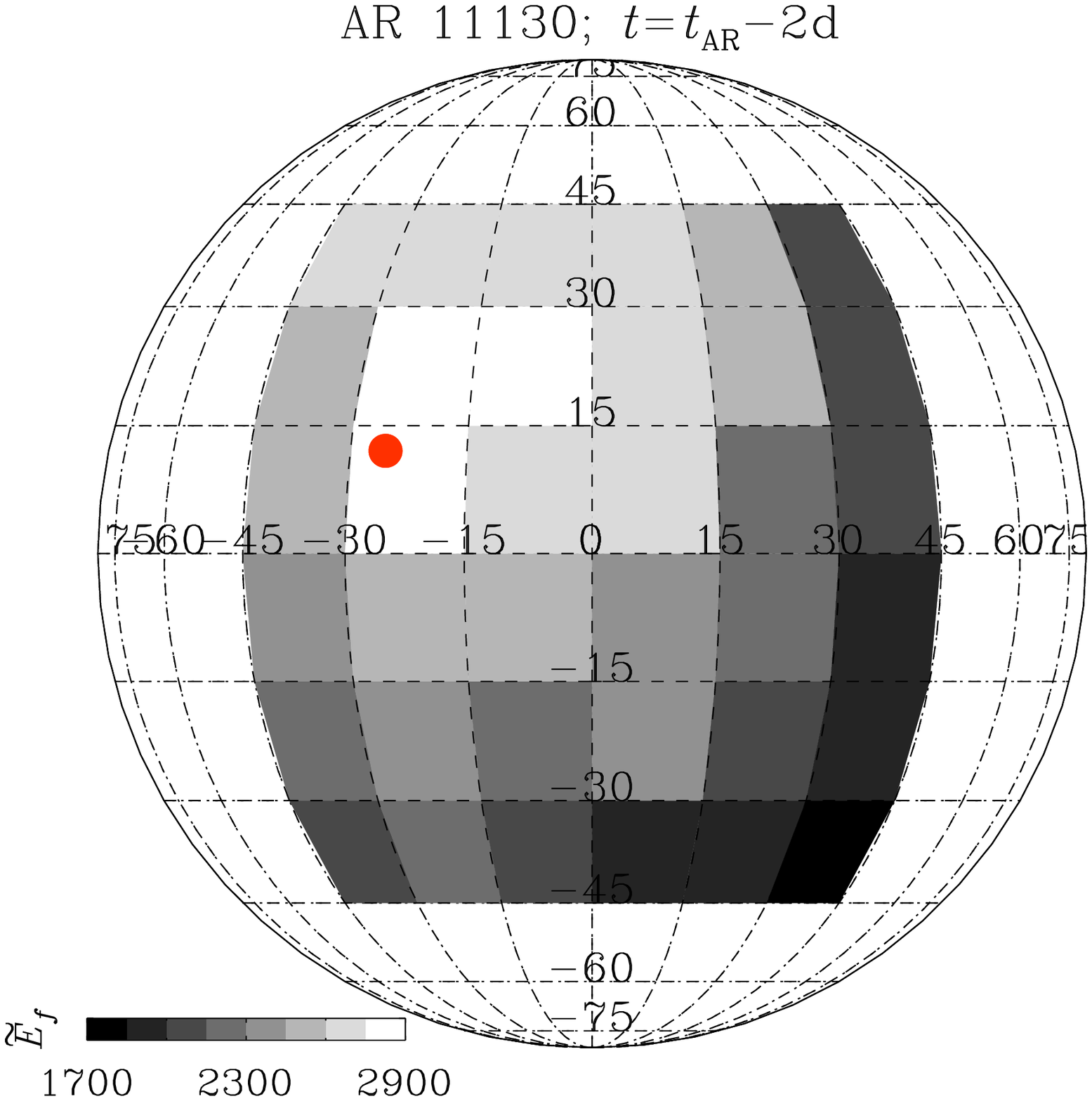}
\vskip0.05in
\includegraphics[width=\columnwidth]{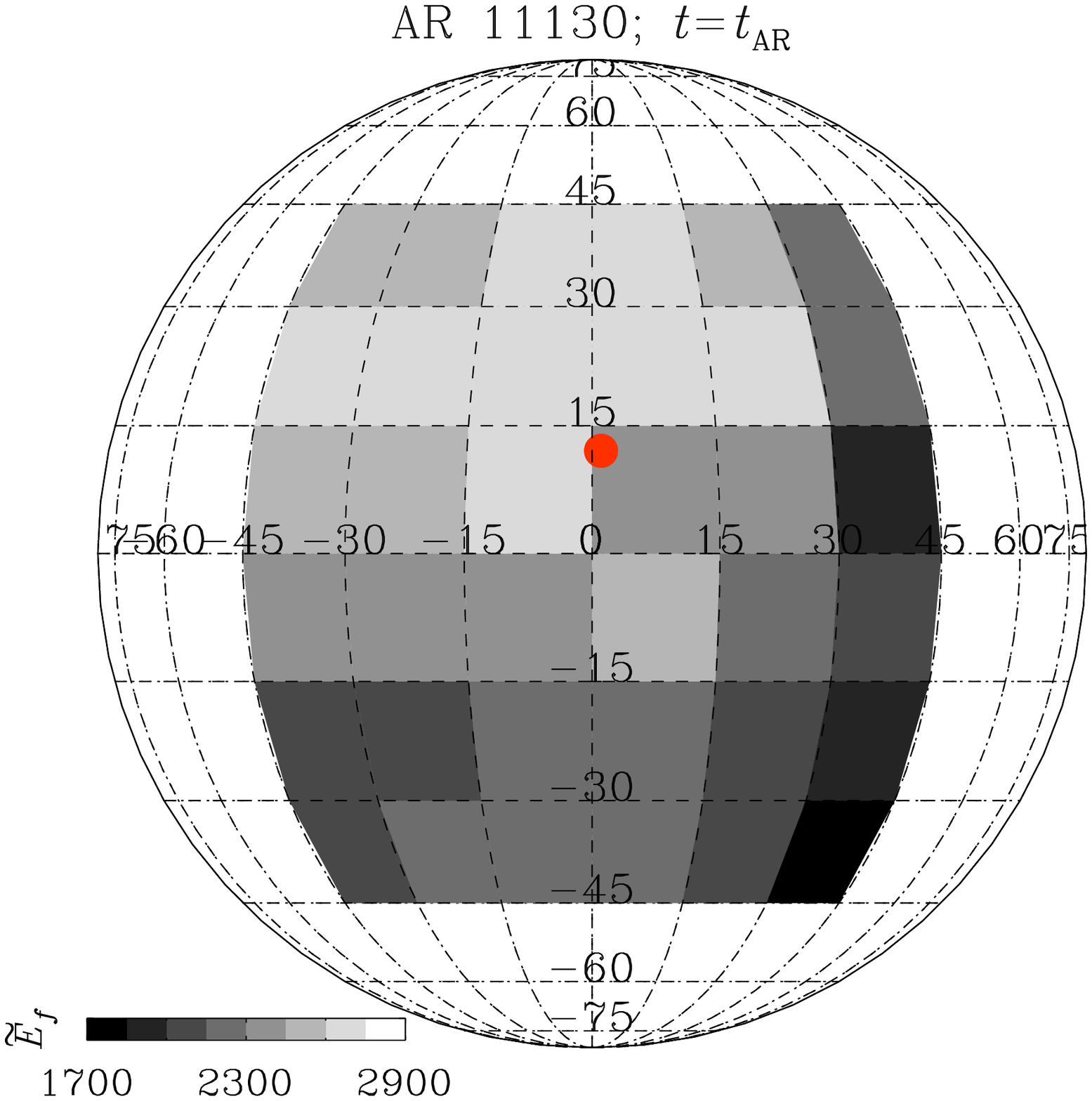}
\end{center}
\caption[]{
Images of $\widetilde{E}_f$ for a relatively quiet phase
of the Sun in 2010 when an isolated AR 11130 emerged on 2010 November 29.
Top and bottom panels show images of
$\widetilde{E}_f$ two days before the AR
emergence and at $t=t_{\rm AR}$, respectively.
Red filled circle denotes the location of AR~11130.
Postel projection mapping was used in constructing these images.
}\label{images}
\end{figure}

\section{Results}
\label{results}

\subsection{Isolated ARs}
\label{iar}

In \Figs{t-trace_QS2010}{t-trace_11130} we compare the quiet sun
during 2010 May 14--19 with an active sun during 2010 November 26--30.
We show the time traces of $\widetilde{E}_f$ for corotating patches.
In \Fig{t-trace_QS2010}, $t_{\rm CM}$ denotes the time of central
meridian crossing while in \Fig{t-trace_11130}, $t_{\rm AR}$ is the
time at which later the AR emerges.
We compare with the time traces of the mirror region
$\widetilde{E}_f^\dag$ in panels (a) of those and subsequent figures,
$\kurt B$ from those patches in panels (b), the rms magnetic
field $\Brms$ within those patches in panels (c), as well as
the corresponding full
disk LoS magnetograms either at $t=t_{\rm AR}-2{\rm d}$
or at $t=t_{\rm AR}-1{\rm d}$ in panels (d)
and at $t=t_{\rm AR}$ in panels (e),
which are the times when the ARs emerge and were assigned their numbers.
The dash-dotted (red) and triple-dot-dashed (orange)
lines in panels (c) denote respectively
the time-traces of $0.08\,B_{\rm max}$ and
$-0.08\,B_{\rm min}$ from the patch of interest where an AR develops.
All six ARs show similar characteristics: an early rise of
$\widetilde{E}_f$ with a maximum $1$--$2$ days prior to $t_{\rm AR}$, followed
by a decline at and after $t_{\rm AR}$, as well as a delayed increase of
$\widetilde{E}_f^\dag$, sometimes with a maximum near $t_{\rm AR}$.
We speculate that the delayed increase of $\widetilde{E}_f^\dag$
might be caused by a correlated response at a distant mirror patch.
This would indicate that the early $f$-mode strengthening, i.e.,
the precursor signal, appears to have an associated causal response at
later times, at distant mirror patches.
Interestingly enough, in most cases, the $\kurt B$ from the patch
where an AR forms also shows a peak before the AR is fully developed,
and thus offers yet another advance indication of AR formation.

By contrast, during a suitably chosen time in 2010, the Sun was nearly
completely quiet, and both $\widetilde{E}_f$ and $\widetilde{E}_f^\dag$
follow each other rather closely (\Fig{t-trace_QS2010}), although their time
traces still show considerable variability.
This might be caused by fluctuations in the subsurface turbulence
and small-scale magnetic fields even for the quiet sun, or perhaps
by instrumental effects.
The fact that $\widetilde{E}_f$
and $\widetilde{E}_f^\dag$ remain close to each other at all times
shows that in the quiet phase of the Sun,
the integrated $f$-mode amplitudes in the two hemispheres
evolve {\em symmetrically}, so that
the difference is small and therefore not significant.
Note also that, since no AR has emerged during that time, we replaced
$t_{\rm AR}$ by the time of central meridian crossing
$t_{\rm CM}$ of an arbitrarily chosen comoving patch
in \Fig{t-trace_QS2010}.

Based on these findings, the following hypotheses may be formulated.
In regions with low or no surface magnetic activity, a nearly flat
time trace without systematic differences between $\widetilde{E}_f$
and $\widetilde{E}_f^\dag$ suggests low subsurface magnetic activity,
while a gradual and systematic enhancement of $\widetilde{E}_f$
relative to $\widetilde{E}_f^\dag$ is suggestive of a build-up of
subsurface magnetic activity.
In already established ARs, on the other hand, $\widetilde{E}_f$
is visibly depressed and $\widetilde{E}_f^\dag$ may or may not show a
marked rise, depending on the complexity of the already established
surface activity.

We adopt a root-mean-square error estimation for
$\widetilde{E}_f$ based on the results shown in \Fig{t-trace_QS2010}
for a magnetically quiet sun. The mean error ($\sigma_{\rm m}$)
is obtained from:
\EQ
\sigma_{\rm m}=\frac{1}{2}\sqrt{\sigma_E^2+
\sigma_{E^\dag}^2}\;;\;
\sigma_E=\sqrt{\left\langle\left(\widetilde{E}_f-
\overline{\widetilde{E}}_f\right)^2\right\rangle}.
\EN
Here, $\langle \widetilde{E}_f \rangle \equiv \overline{\widetilde{E}}_f$
denotes the mean value of $\widetilde{E}_f(t)$.
We use $\sigma_{\rm m}$ to display error bars in figures showing
$\widetilde{E}_f(t)$.

\begin{figure}
\begin{center}
\includegraphics[width=\columnwidth]{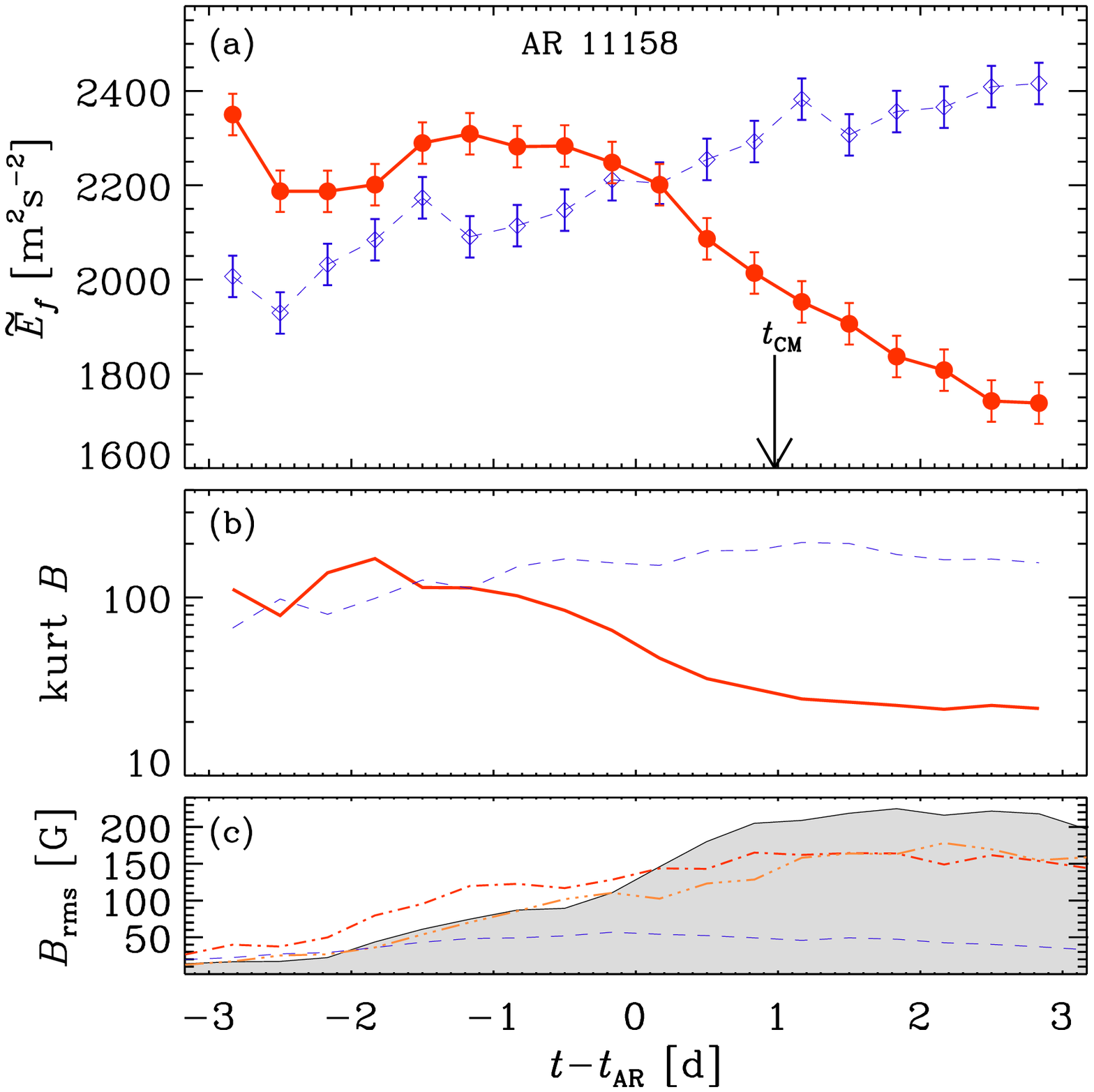}
\includegraphics[width=\columnwidth]{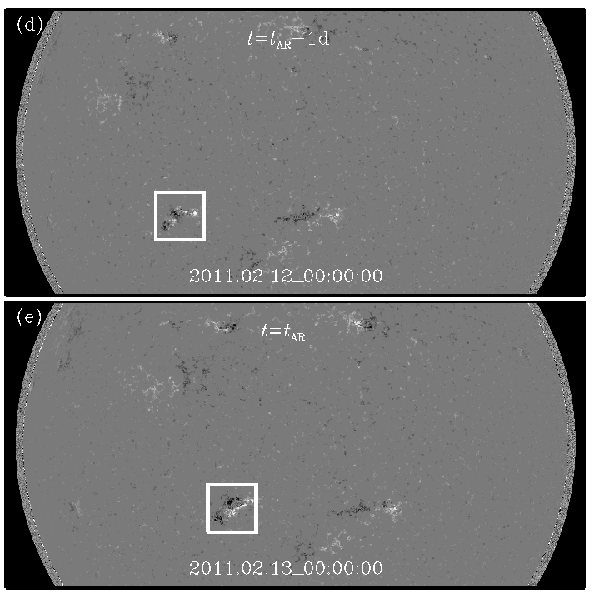}
\end{center}
\caption[]{
Same as \Fig{t-trace_11130}, but for AR~11158.
}\label{11158}
\end{figure}

\begin{figure}
\begin{center}
\includegraphics[width=\columnwidth]{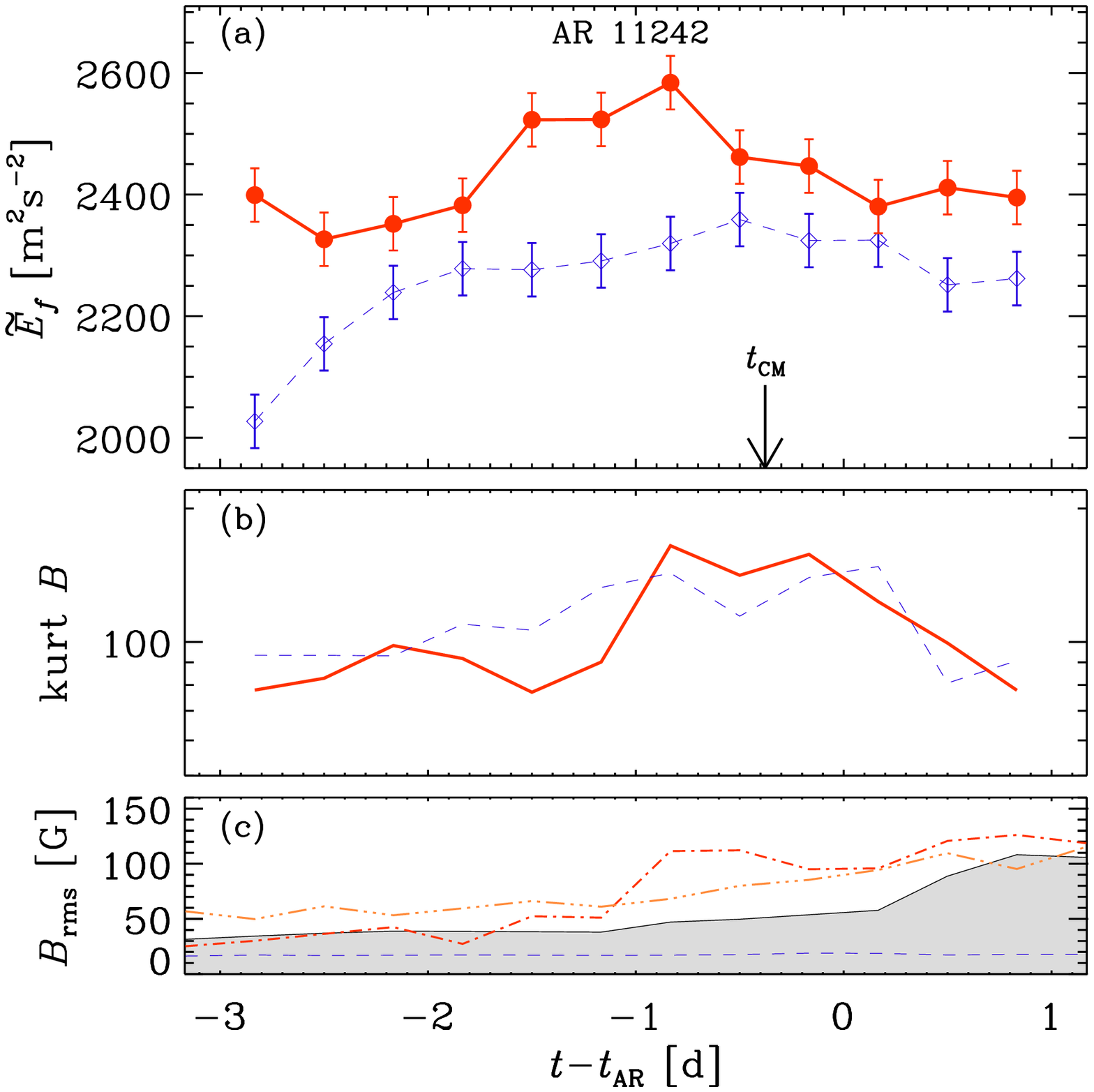}
\includegraphics[width=\columnwidth]{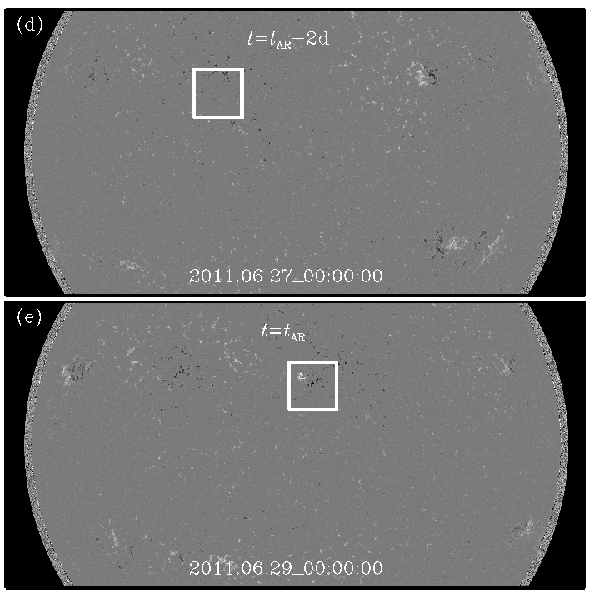}
\end{center}
\caption[]{
Same as \Fig{t-trace_11130}, but for AR~11242.
}\label{11242}
\end{figure}

\begin{figure}
\begin{center}
\includegraphics[width=\columnwidth]{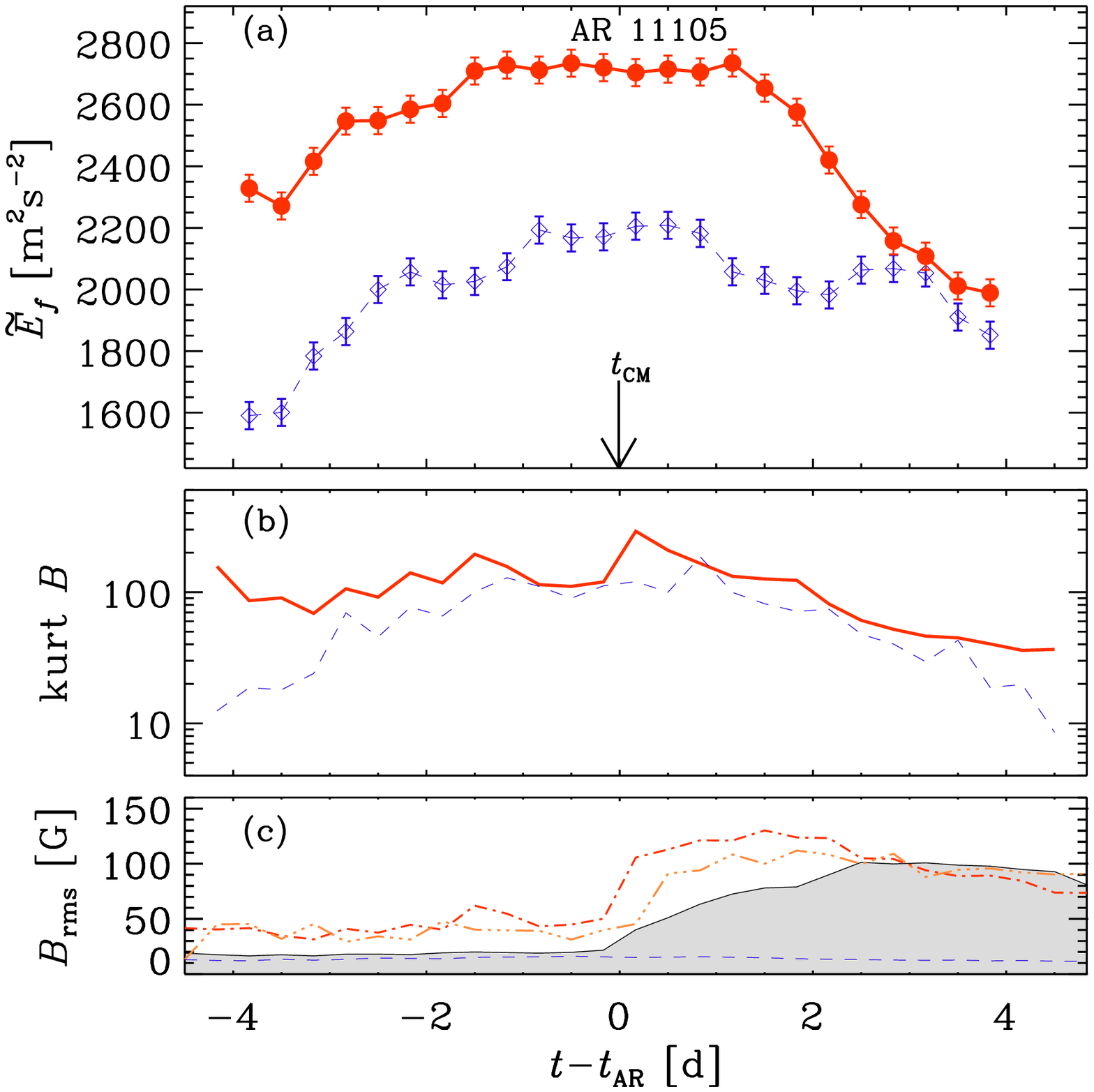}
\includegraphics[width=\columnwidth]{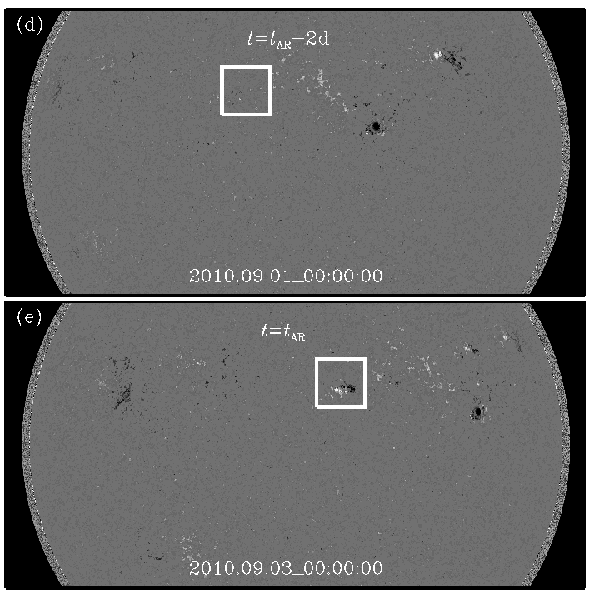}
\end{center}
\caption[]{
Same as \Fig{t-trace_11130}, but for AR~11105.
}\label{11105}
\end{figure}

\begin{figure}
\begin{center}
\includegraphics[width=\columnwidth]{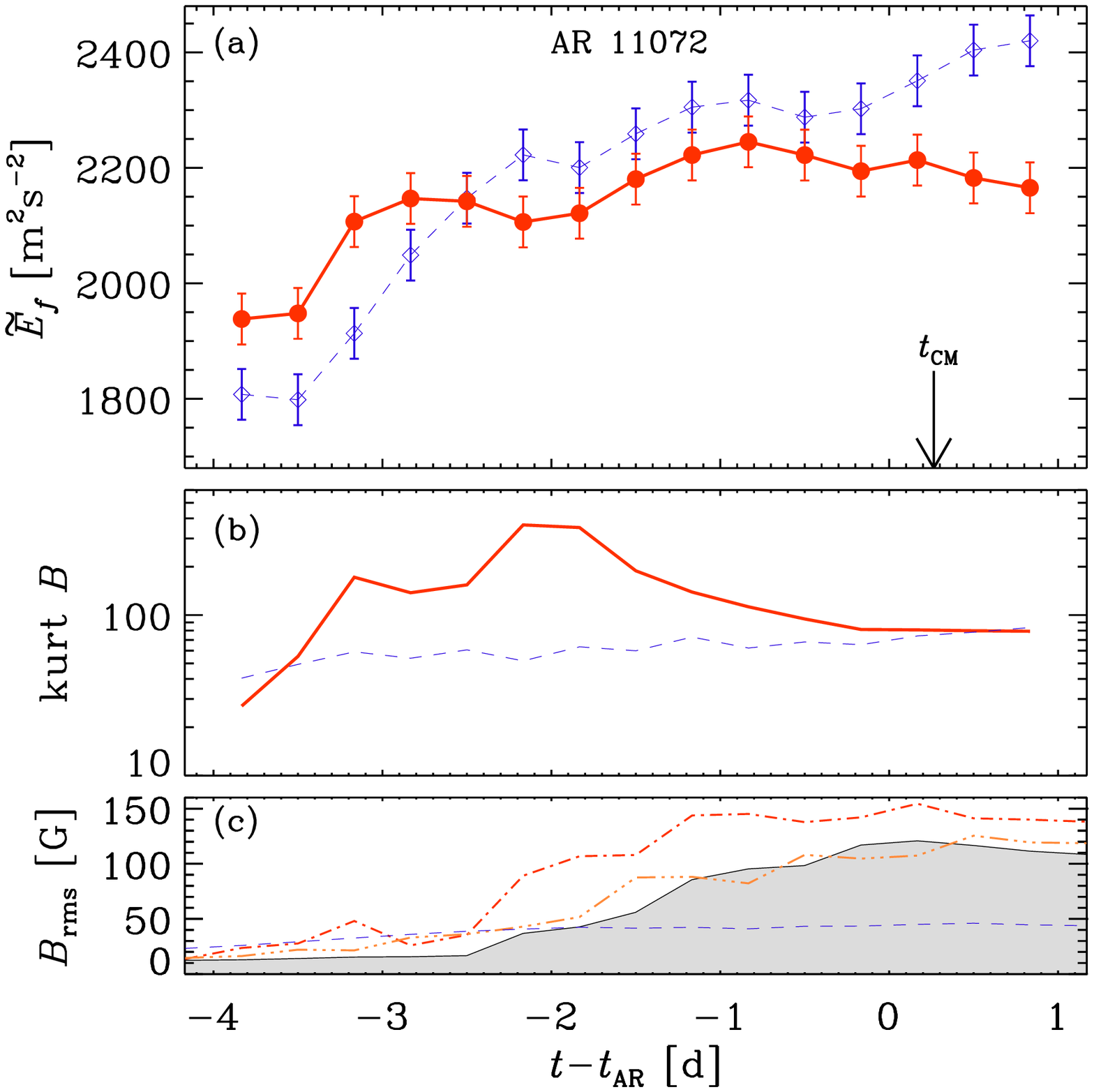}
\includegraphics[width=\columnwidth]{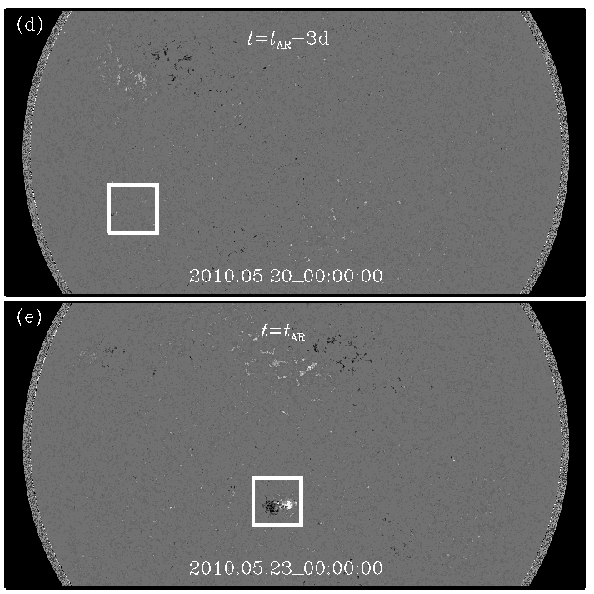}
\end{center}
\caption[]{
Same as \Fig{t-trace_11130}, but for AR~11072.
}\label{11072}
\end{figure}

\begin{figure}
\begin{center}
\includegraphics[width=\columnwidth]{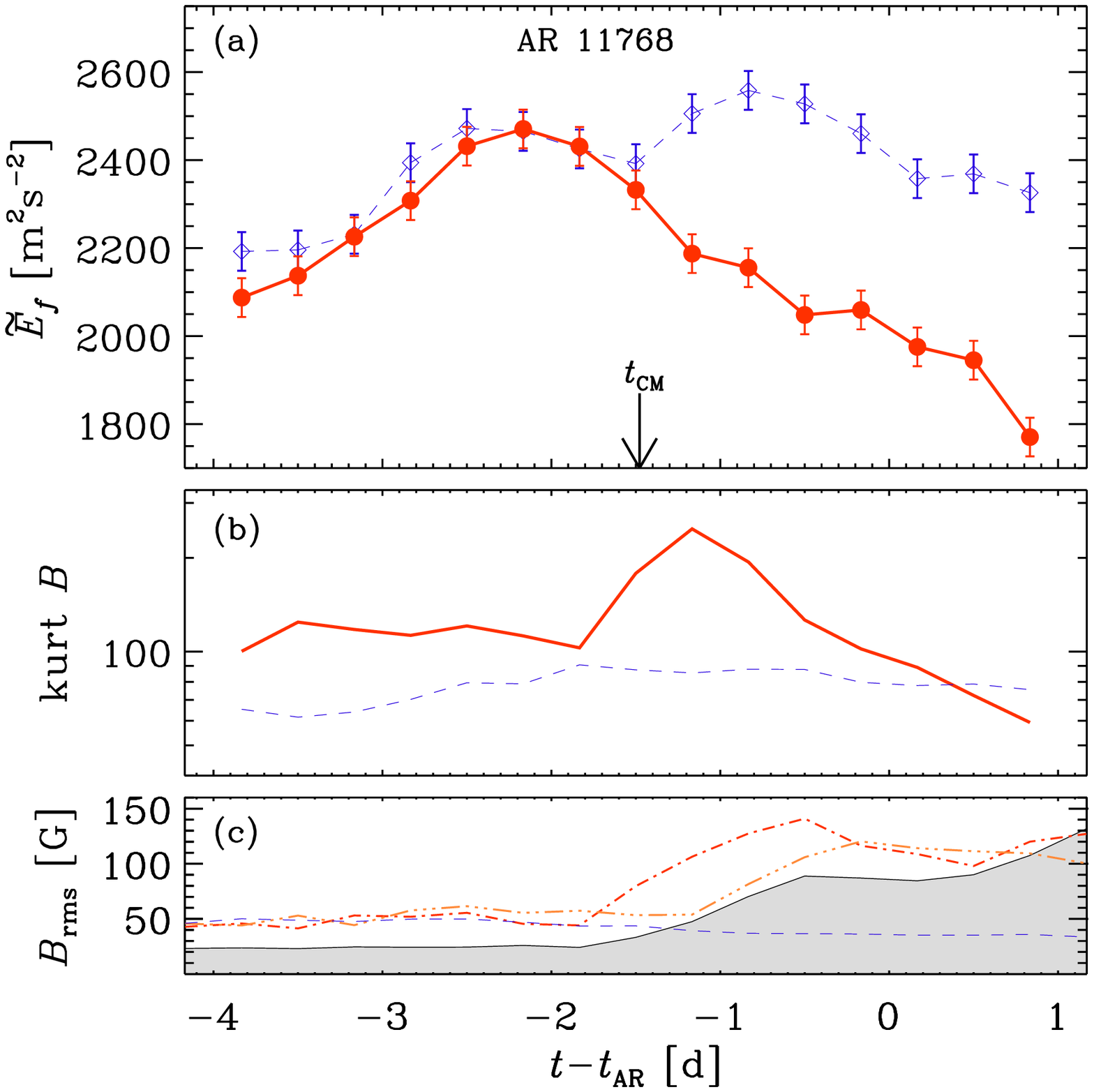}
\includegraphics[width=\columnwidth]{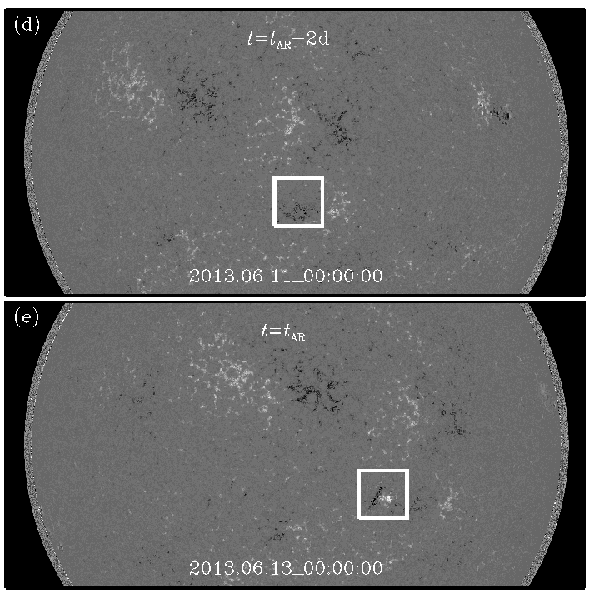}
\end{center}
\caption[]{
Similar to \Fig{t-trace_11130}, but for AR~11768.
}\label{11768}
\end{figure}

\begin{figure}
\begin{center}
\includegraphics[width=\columnwidth]{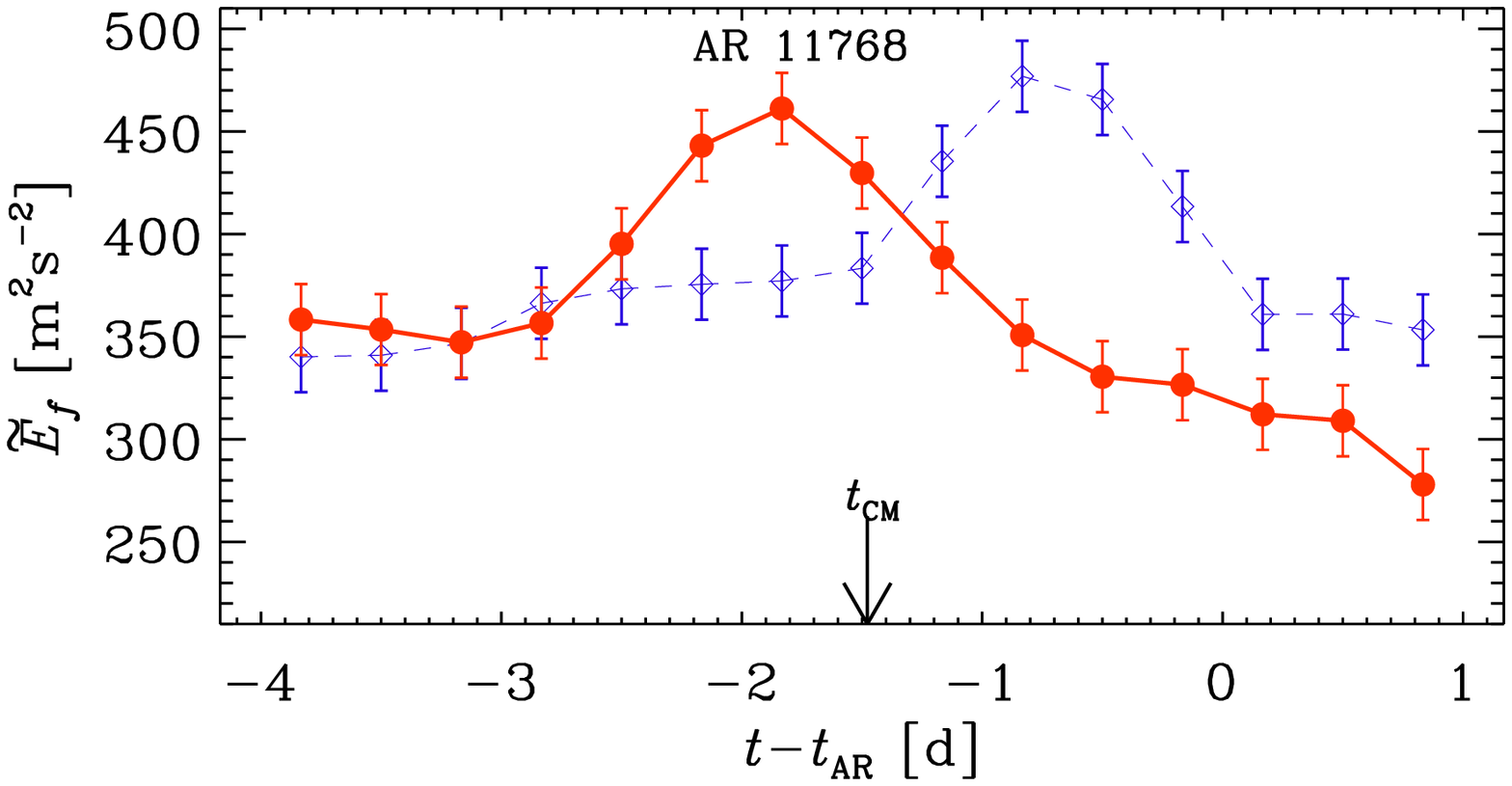}
\end{center}
\caption[]{
Similar to \Figp{11768}{a}, but using $k_y R_\odot \in [1200,1300]$
instead of $[1200,2000]$ in determining $E_f$ using \Eq{ef}.
}\label{11768_a2}
\end{figure}

Let us now discuss the individual examples in more detail.
AR~11130 was a solitary AR during 2010 when
the overall solar activity was still rather low.
It is therefore an example where interference from other locations
on the Sun is minimal.
Indeed, it displays most strikingly the ``symmetry breaking'' between
$\widetilde{E}_f$ and $\widetilde{E}_f^\dag$, with $\widetilde{E}_f$
showing a maximum about 1.5 days before this AR emerges;
see \Fig{t-trace_11130}.
Also, $\kurt B$ shows a peak more than one day before this AR is
fully developed; see the solid red line in \Figp{t-trace_11130}{b}.

As an extension of this work, we also calculate
images of $\widetilde{E}_f$ for the solar disk.
This gives more explicit information of where the next AR might form;
see \Fig{images} showing images at times when AR~11130 was forming.
It is remarkable that the maximum in $\widetilde{E}_f$ at time
$t=t_{\rm AR}-2{\rm d}$ coincides with the location (marked by
a red filled circle) where AR~11130 is going to form later. 
We also note from the top image in \Fig{images} that the strengthening
of the $f$-mode about two days prior to the emergence of the
AR is nonlocal in space, with patches progressively farther
from the predicted location of AR~11130 showing almost monotonically
decreasing $\widetilde{E}_f$.
Although we see a moderate degree of fluctuation, there are also
systematic effects---especially near the limb.
Whether or not these are caused by instrumental effects such as variations
of the modulation transfer function \citep{Wachter} is unclear.
If so, the remaining variations may either also be related to instrumental
effects or they could be caused by weaker subsurface magnetic fields
that must always be present---even during solar minimum.

Next, we consider AR~11158 (\Fig{11158}), which was a rapidly growing AR
that produced the first X-class flare of solar cycle 24 on 2011 February~15
\citep{MVA12} with an Earth-directed halo coronal mass ejection \citep{Sch11}.
It also produced several M-class flares during February 13--16 \citep{Ino13},
after being assigned its number on February~13.
Also in this case, $\widetilde{E}_f$ shows a clear increase with
$\widetilde{E}_f-\widetilde{E}_f^\dag\sim200\m^2\s^{-2}$ about
a day before $\Brms$ reaches a plateau of about $220\G$.
The energy increase of about $\sim300\m^2\s^{-2}$ seen about
three days prior to the AR emergence appears to be indicative of a
subsurface concentration of the magnetic field resulting in a
rapid growth of $\Brms$ in the photosphere.
Thus, the same general trend is found here too, although
the potential for using $\widetilde{E}_f(t)$ as a precursor was
less clear in the sense that it showed a maximum only about a day in
advance.
The subsequent increase in $\widetilde{E}_f^\dag$
is noticeable here as well.
In this case, $\kurt B$ shows a peak already at $t=t_{\rm AR}-2{\rm d}$.

AR~11242 (\Fig{11242}) was assigned its NOAA number on 2011 June 29,
a day before it fully emerged in isolation.
Here we find elevated values of $\widetilde{E}_f$ relative to
$\widetilde{E}_f^\dag$ for all times during our tracking period,
where $\widetilde{E}_f$ shows a maximum about 1--2 days prior to
AR formation.
Again, early strengthening of $\widetilde{E}_f$
at $t=t_{\rm AR}-3{\rm d}$ appears as a precursor to the
rise of $B_{\rm max}$ and the peak in $\kurt B$ at
$t=t_{\rm AR}-1{\rm d}$. 
In this case too, we have strong evidence of $f$-mode strengthening
about 1--3 days before there is any visible magnetic activity
at the patch where AR~11242 develops later.
\cite{Smi13} reported long-period
oscillations of 200--400 min
associated with this AR, using simultaneous
data from HMI and ground-based radio emission measurements at 37 GHz
from Mets\"ahovi radio observatory at Aalto University in Finland.
They interpreted their results based on the shallow sunspot model
of \cite{SK09}, which may even show some resemblance to the magnetic
flux concentrations that form spontaneously in strongly stratified
turbulence simulations \citep{BKR13}.

Now we consider the case of AR~11105 (\Fig{11105}), which was assigned its
NOAA number nearly at the time of onset of $\Brms$ on 2010 September 3.
For AR~11105, similar to the previous example, $\widetilde{E}_f$ 
remains larger than $\widetilde{E}_f^\dag$ during the tracking,
and shows the usual post-emergence damping.
Unlike the other examples, the time trace of $\kurt B$
is in this case featureless.

Next we turn to AR~11072 (\Fig{11072}), which was identified on 2010 May 23
when $\Brms$ had reached its peak value, although $B_{\max}$ from
the same region showed an early growth already about two days earlier.
About four days prior to $t_{\rm AR}$,
the patch in this case was much closer to the limb than in the
other examples and the data might have suffered some systematic effects,
as discussed above. However, we do find weak signatures of relative
strengthening of $\widetilde{E}_f$ at
$t\approx t_{\rm AR}-3.5{\rm d}$, although the damping of the $f$-mode
after the flux emergence is not seen. This might be due to the
episodic flux emergences in this case, as is apparent from
\Fig{11072}. Interestingly, $\kurt B$ shows a sharp rise at about
the same time when we find signs of $f$-mode strengthening,
and it exhibits a double-peaked feature, which is all much before $\Brms$
saturates in this region.

For AR~11768, we now perform the following experiment to highlight the
significance of wavenumber dependence of the proposed precursor
signal, i.e., the $f$-mode strengthening, and have presented our
results for this case in \Figs{11768}{11768_a2}.
We considered two different wavenumber intervals in determining
$E_f$ using \Eq{ef}; while \Figp{11768}{a} corresponds to the same range,
$k_y R_\odot \in [1200,2000]$, as used in the other cases, for
\Fig{11768_a2} we chose a much narrower wavenumber range,
$k_y R_\odot \in [1200,1300]$, which explains the lower
values of $\widetilde{E}_f$.
As shown in \Fig{11768_a2}, there is again the
characteristic symmetry breaking between $\widetilde{E}_f$
and $\widetilde{E}_f^\dag$, with $\widetilde{E}_f$ showing
a maximum at $t=t_{\rm AR}-2{\rm d}$. In this case
the initial rise is sharper than, say, for AR~11130.
However, no such relative strengthening of $\widetilde{E}_f$
is seen before emergence when the larger wavenumber range is considered,
although the usual rise of $\widetilde{E}_f$ followed by
the post-emergence damping is clearly visible; see
\Figp{11768}{a}.
This provides a hint of a possible wavenumber dependence of the effect
causing the $f$-mode strengthening prior to AR formation.
The perturbed wavenumbers of the $f$-mode correspond to horizontal
scales of around $3000\kmeter$ and we speculate that these might be
the typical scales of magnetic structures that are gradually growing
in both strength and size while retaining their imprints
in terms of causing the observed $f$-mode strengthening correspondingly
at such high wavenumbers.
Note that in this case, large-scale patches of weak magnetic
fields are present in the opposite hemisphere,
with $\Brms^\dag \approx 2\Brms$ at early times, being highest of
all the other cases. This could affect the values of $\widetilde{E}_f^\dag$.
Here too, $\kurt B$ shows a peak about a day before the magnetic
flux associated with this AR is fully emerged.

\begin{figure}
\begin{center}
\includegraphics[width=\columnwidth]{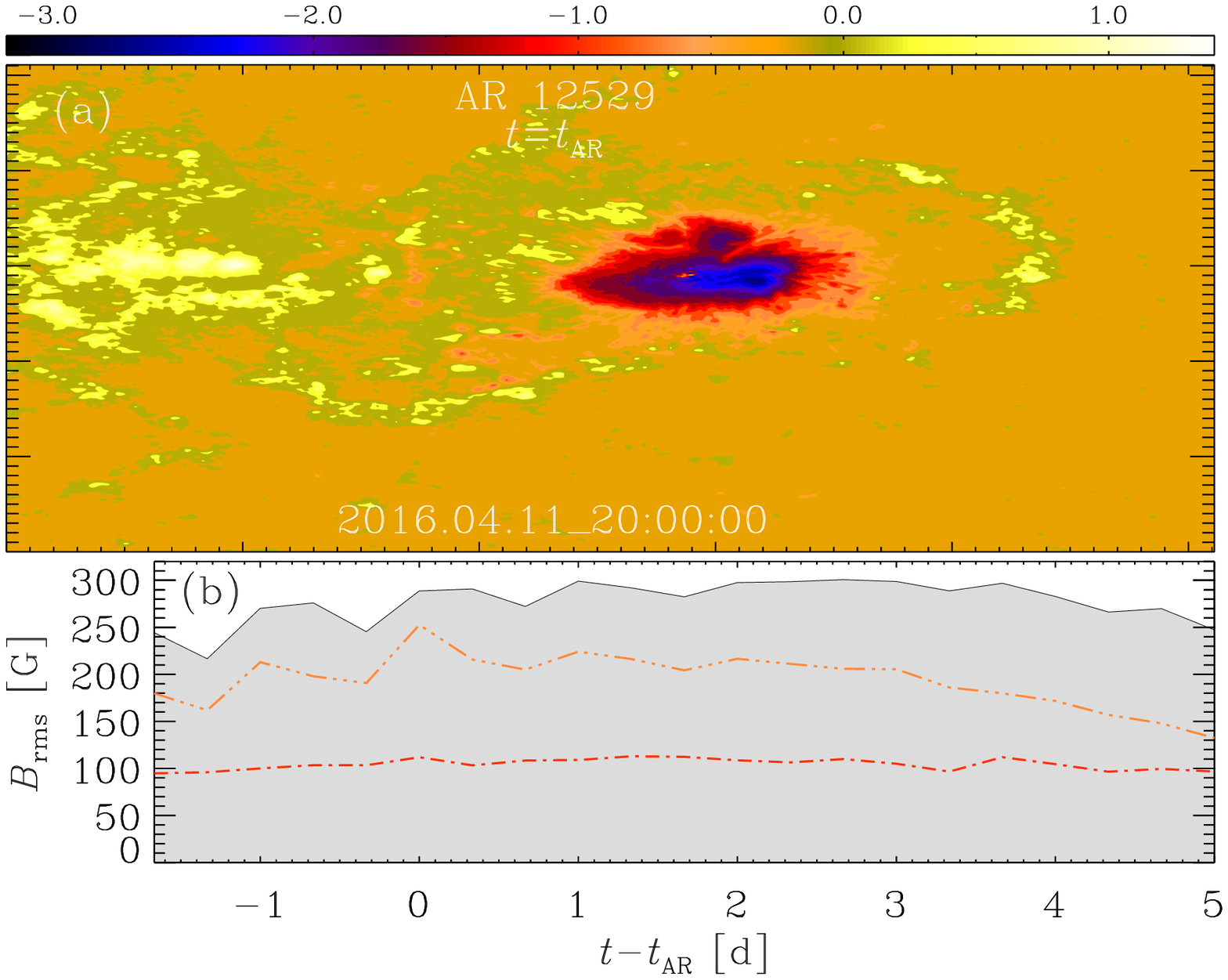}
\end{center}
\caption[]{
Panel (a) shows the AR~12529 with
colors indicating the strength of the LoS magnetic field $B$ in
$\kG$; panel (b) shows temporal evolution of its rms
strength $\Brms$ (solid line with shaded area underneath)
together with $\Brms^\dag$ (dashed blue line).
The dash-dotted (red) and triple-dot-dashed (orange)
lines denote time-traces of $0.08\,B_{\rm max}$ and $-0.08\,B_{\rm min}$,
respectively from the patch shown in panel (a).
}\label{QSnear12529_fg}
\end{figure}

\begin{figure}
\begin{center}
\includegraphics[width=\columnwidth]{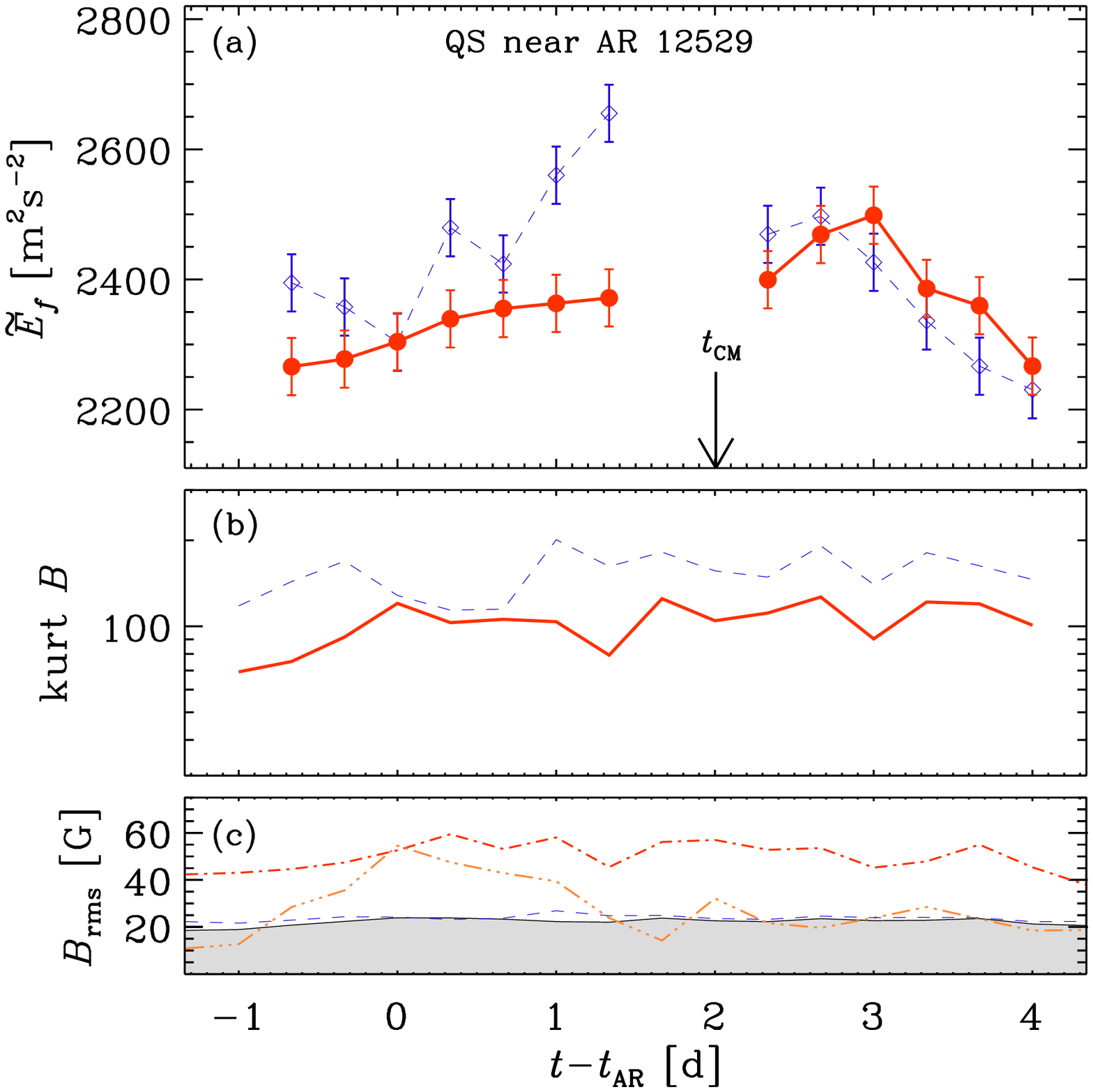}
\includegraphics[width=\columnwidth]{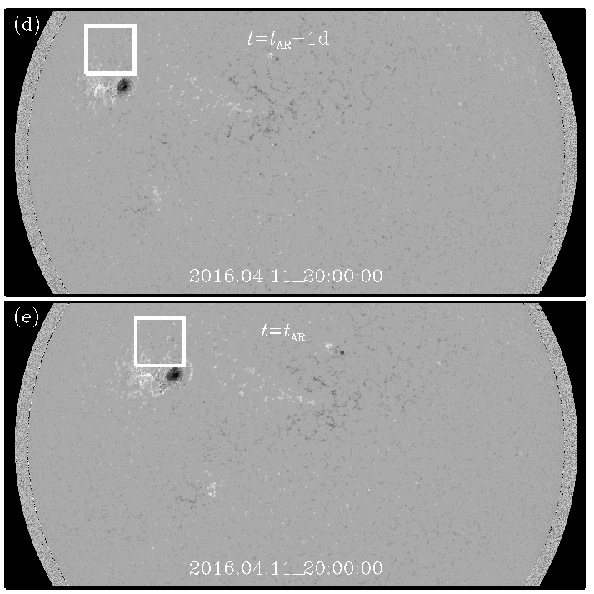}
\end{center}
\caption[]{
Similar to \Fig{t-trace_11130}, but for a magnetically quiet
patch lying next to AR~12529.
Here, $t=t_{\rm AR}$ corresponds to a maximum in $|B_{\rm min}|$.
}\label{QSnear12529}
\end{figure}

\begin{figure}
\begin{center}
\includegraphics[width=\columnwidth]{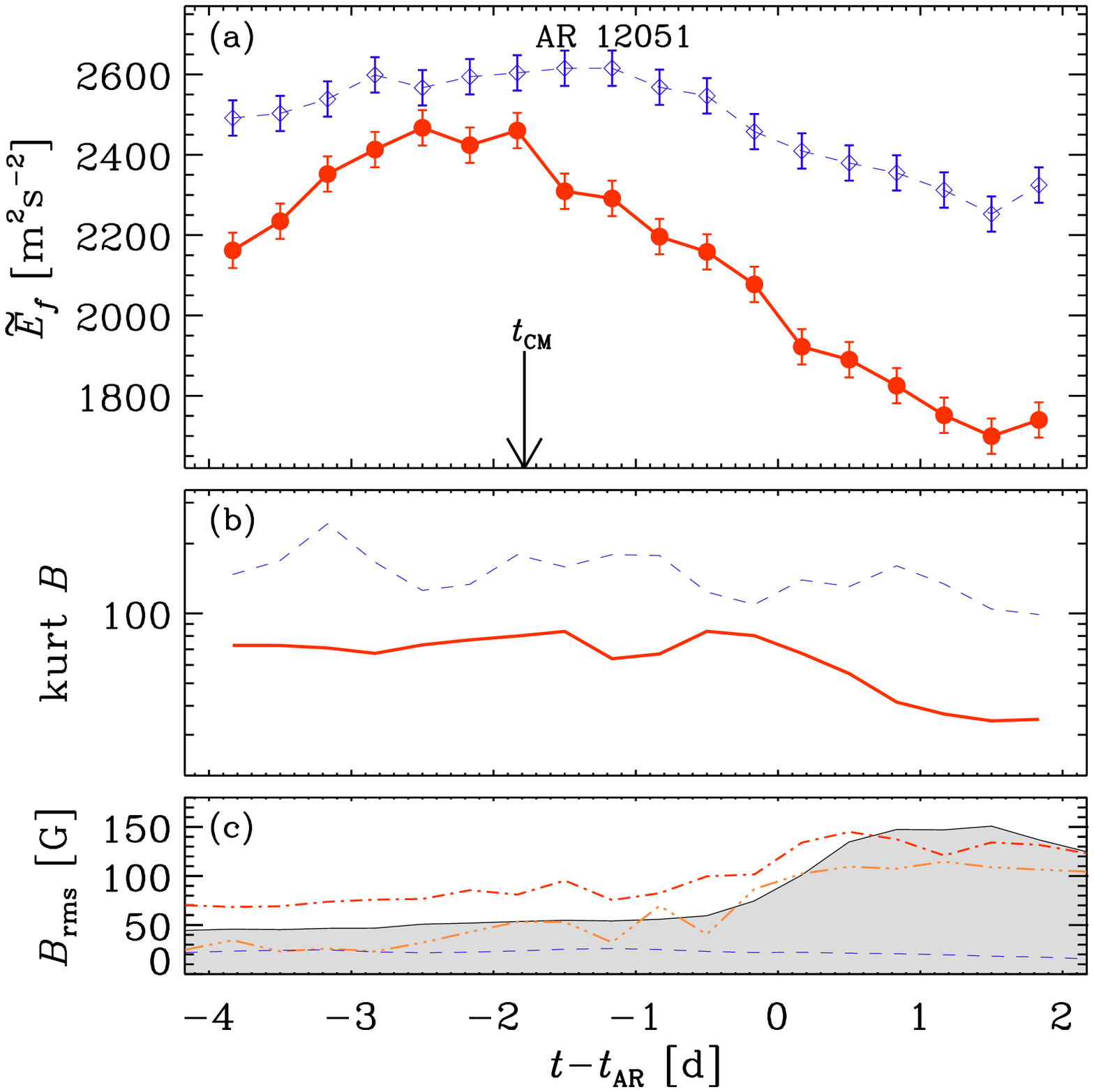}
\includegraphics[width=\columnwidth]{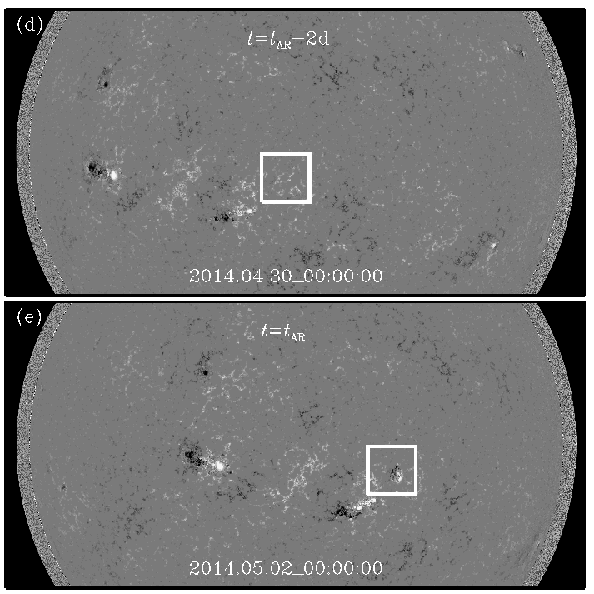}
\end{center}
\caption[]{
Same as \Fig{t-trace_11130}, but for AR~12051,
which is in close proximity to already existing ARs.
}\label{12051}
\end{figure}

Comparing now all the six ARs in our sample, we see that the three ARs
that appeared in the north (ARs~11130, 11242, and 11105) had slightly
larger values of $\widetilde{E}_f$ (2600--2800~${\rm m^2s^{-2}}$)
than the three in the south (ARs~11158, 11072, and 11768 with
$\widetilde{E}_f$=(2200--2400~${\rm m^2s^{-2}}$).
This is consistent with the strong north-south asymmetry of cycle~24
with stronger activity and an earlier maximum in the north and weaker
activity and a later maximum in the south \citep{CCG13,SHLZ15}.
This shows that the value of $\widetilde{E}_f$ reflects the general
subsurface magnetic activity even over the time scale of the solar cycle.

\subsection{Crowded ARs}
\label{car}

As argued above, the ARs cause damping of the $f$-mode
after their emergence and this might influence the signal
from a newly forming AR in the neighborhood.
In order to extract precursor signals from a developing AR in a crowded
environment, we perform an experiment demonstrating the nonlocality
of the high-wavenumber $f$-mode damping caused by already established
ARs. Here, we consider a magnetically quiet patch lying just
above AR~12529, which was an already existing strong
AR during 2016 April; see \Figp{QSnear12529_fg}{a}, which shows a close-up
of this AR and the temporal evolution of its $\Brms$ in panel (b).
Similar to, say, \Fig{t-trace_11130}, we show time traces of
$\widetilde{E}_f$ (corresponding to the quiet patch above AR~12529)
and $\widetilde{E}_f^\dag$ in \Fig{QSnear12529}.
Note that both patches being tracked in this experiment are
magnetically quiet and that their corresponding kurtoses are
essentially featureless. We find a significant damping of
$\widetilde{E}_f$ as compared to $\widetilde{E}_f^\dag$ at
$t\approx t_{\rm AR}+1{\rm d}$, after which there are some
data gaps in the observations. Both $\widetilde{E}_f$ and
$\widetilde{E}_f^\dag$ attain similar values at late stages.

We now turn to the case of AR~12051, which lies next to
bigger and stronger ARs that had appeared already in the
southern hemisphere; see the magnetograms in
panels (d) and (e) of \Fig{12051}.
Here too we find that the evolution of $\widetilde{E}_f$
obtained from the patch where later AR~12051 emerges
is not flat; see \Fig{12051}.
It rises from a level of about $\sim$2150 ${\rm m^2s^{-2}}$ and
attains a maximum of $\sim$2400 ${\rm m^2s^{-2}}$
more than two days before it was assigned its number on 2014 May 2 and
nearly three days before $\Brms$ reached its maximum value of about $150\G$.
On May 3, this AR developed a so-called $\delta$-class spot with M class
flares a few days later.
However, the essential difference
here is that the $\widetilde{E}_f^\dag$ from the relatively quiet
mirror patch remains larger than $\widetilde{E}_f$ at all times.
This might well be expected based on our experiments and results
presented above and may be understood as follows:
as the southern hemisphere is already
``polluted'' by many ARs, the $f$-mode is expected to be damped in this
hemisphere and therefore the time-trace of $\widetilde{E}_f$ for
AR~12051, while showing early precursor signatures, does not overcome
$\widetilde{E}_f^\dag$ from the northern hemisphere where the
$f$-mode remains undamped and shows a much smaller variation.

\begin{figure}
\begin{center}
\includegraphics[width=\columnwidth]{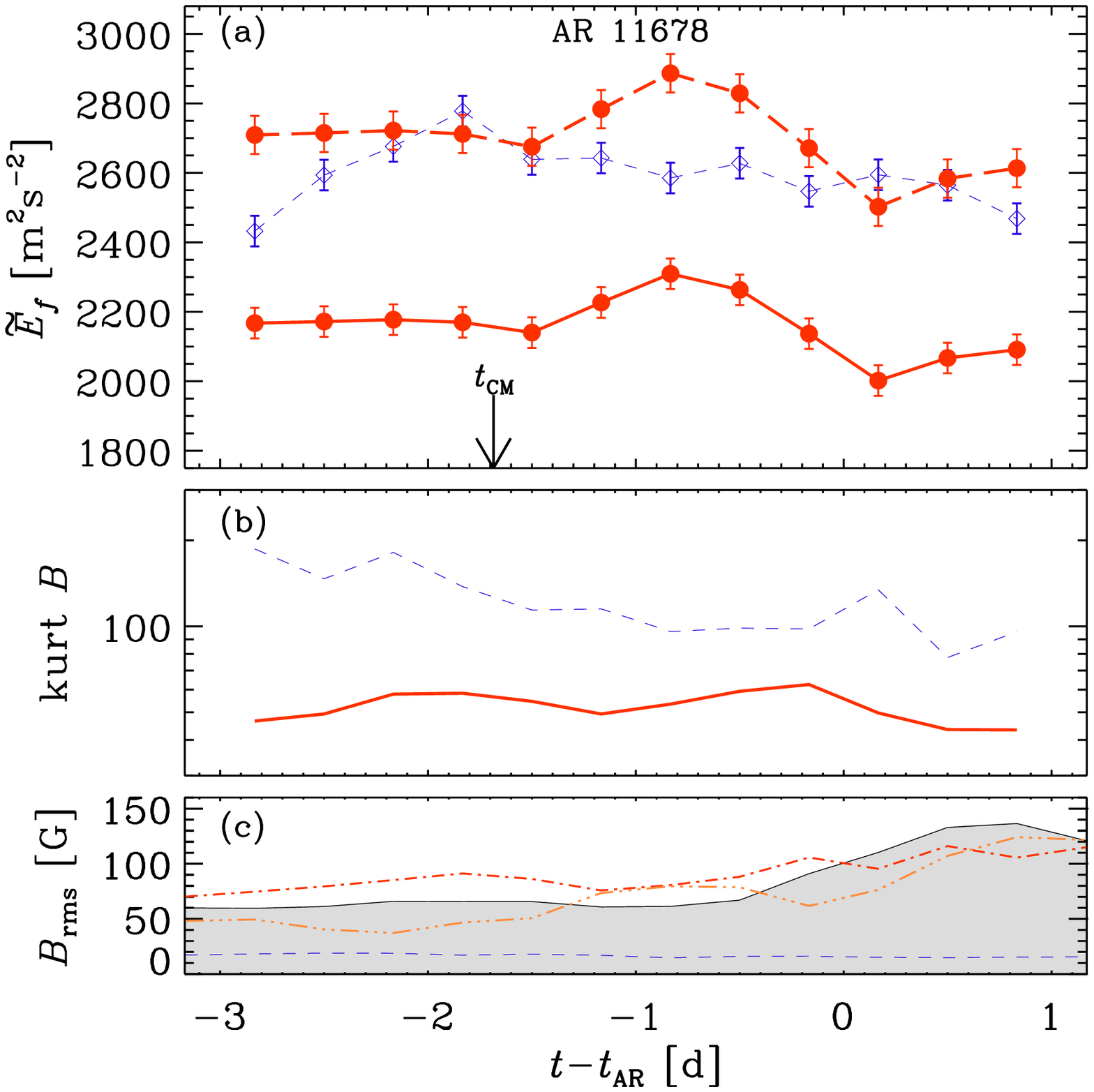}
\includegraphics[width=\columnwidth]{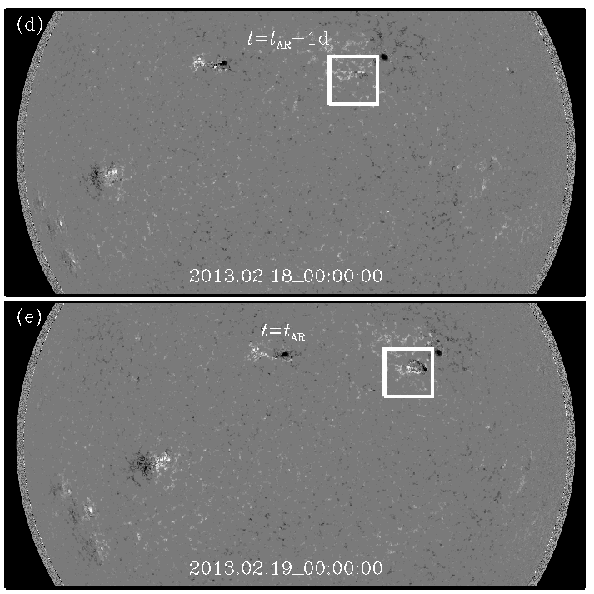}
\end{center}
\caption[]{
Same as \Fig{12051}, but for AR~11678. Here, the dashed
(red) line in panel (a) corresponds to $1.25\widetilde{E}_f$.
}\label{11678}
\end{figure}

Having discussed the possible difficulties in predicting a new
AR emerging in a crowded environment, we now wish to describe
plausible procedures that might be useful in still extracting the
precursor signals in such ``polluted'' medium.
One may be able to find guidance
from a standard technique of optical astronomy where one
routinely subtracts emission from a bright foreground star in order
to detect and study a faint background source.
In the present context, this would require a more detailed knowledge
of the $f$-mode damping mechanism caused by existing ARs on the
solar disk, so that one could apply a similar \emph{cleaning}
procedure. We make such an attempt for our final case of
AR~11678, which emerged next to a group of compact ARs on
2013 February 19. In \Fig{11678} we show such a plot.
Although the time-trace of $\widetilde{E}_f$ shows a peak about
a day before this AR emerges, it remains smaller than
$\widetilde{E}_f^\dag$ at all times of tracking,
as would be expected in this case. Based on the other
cases discussed earlier, we find that the amount of observed
damping of $\widetilde{E}_f$ could be as large as about
$25\%$ of its peak value. Therefore, we applied a uniform boost
of $25\%$ to the original $\widetilde{E}_f$ in an attempt to
correct against the expected damping, and show the thus boosted
time-trace of $1.25\widetilde{E}_f$ by dashed red line with filled
circles in \Figp{11678}{a}.
This immediately reveals the relative
strengthening---similar to what is observed in the case of isolated ARs.
This experiment with a uniform boost is meant to
highlight the necessary correction procedure.
Clearly, we need better knowledge of the post-emergence effects on
the $f$-mode, not only locally but also in the surrounding medium,
to be able to apply a realistic, non-uniform boost that depends on
the magnetic activity in the neighborhood.

\section{Implications}

If we accept that $\widetilde{E}_f(t)$ can be used as a
precursor to AR formation,
we must ask about its possible physical origin and relevance.
Earlier idealized simulations \citep{SBR14,SBCR15} have demonstrated that,
while uniform magnetic fields lead to a frequency shift and a weakening
of the $f$-mode, a nonuniform subsurface field can lead to a fanning and
associated strengthening of the $f$-mode, provided the magnetic field
is at least one or two pressure scale heights below the surface.
While such studies should be repeated with more realistic models,
they do confront us with the question of how a magnetic field can remain
undetected once it is only a few Mm below the surface.

The fact that the $f$-mode resides near the top few $\Mm$ of the Sun
is suggestive of a gradual build-up of the AR near the surface, instead of
a buoyant rise, which would happen in just a few hours \citep{CRTS10}.
This is in stark contrast to the conventional picture of an $\Omega$-shaped
flux tube rising from the bottom of the convection zone and forming an AR
as it pierces the surface \citep{Fan01}.
Earlier simulations of \cite{CRTS10} with a magnetic field implanted
at a depth of nearly $10\Mm$ below the surface have produced surface
manifestations just a few hours later.
Such simulations do not address the physics
of the {\em formation} of magnetic flux concentrations.
By contrast, several simulations in large enough domains performed
by several groups \citep{SN12,WLBKR13,MBKR14,KBKKR16,MS16} have
demonstrated the spontaneous emergence of magnetic flux concentrations
right at the surface.
This highlights the potential significance of $f$-mode-related precursors
at constraining our still very sketchy understanding of the solar
dynamo \citep{Ossen03,B05,Cha10}.
Yet another important quantity to be investigated in dynamo models is the
kurtosis of the magnetic field. Models exhibiting a peak in
$\kurt B$ well before $\Brms$ saturates are expected to be better
constrained and might become more favorable.

\section{Conclusions}

All six examples of isolated ARs presented in \Sec{iar} show
that, several days prior to magnetic field emergence, the
strength of the $f$-mode, as presented by
the value of $\widetilde{E}_f$, rises and then reaches
a maximum before displaying the known post-emergence damping.
Also, prior to AR emergence, the value of $\widetilde{E}_f$
remains larger for long times
with significant energy difference compared to the value
obtained from the corresponding quiet sun location,
$(\vartheta^\dag, \varphi)$.
For the two examples of crowded ARs presented in \Sec{car}, however,
this is different and, as explained above, the reason
for this is in fact expected.
We summarise our findings as follows:
\begin{itemize}
\item The solar $f$-mode is perturbed and shows a strengthening at
high wavenumbers caused by the subsurface magnetic fields
associated with emerging ARs about 1--2 days before there is any visible
magnetic activity in the photosphere.
This appears to be independent of the phase within the solar cycle.
\item We discussed the wavenumber dependence of the precursor signal
and showed that the $f$-mode strengthening occurs at fairly large
wavenumbers.
\item In many cases, the kurtosis of the magnetic field from the patch
in which the AR develops shows a peak much before $\Brms$ from that
region saturates.
\item As discussed in earlier works, we find that the $f$-mode suffers
damping after the emergence of the AR.
\item The $f$-mode strengthening prior to AR formation, followed
by its post-emergence damping, are nonlocal in space, and thus could
influence the neighboring patches.
\item We proposed a plausible cleaning procedure to extract precursor
signal from patches in a crowded environment with one or more
pre-existing ARs.
\end{itemize}

Calculating images of $\widetilde{E}_f$ for the solar disk,
as shown by an example in \Fig{images}, appears to provide
explicit information of where the next AR might form. But
we need more studies to better understand the post-emergence damping
of the $f$-mode and its effects on the surrounding medium
in order to calibrate the necessary correction/cleaning
that must be applied to the data to extract precursor signals
from a polluted medium.

\acknowledgments

We thank Charles Baldner, Aaron Birch, Rick Bogart, Robert Cameron,
Brad Hindman, Maarit K\"apyl\"a, Charlie Lindsey,
Matthias Rheinhardt, Jesper Schou, Hannah Schunker, Sami Solanki,
Junwei Zhao, and the referee for their comments and suggestions.
This work has been supported in parts by
the Swedish Research Council grant No.\ 621-2011-5076
as well as a startup grant from CU-Boulder.



\begin{thebibliography}{99}

\bibitem[Birch et al.(2010)]{BBF10}
Birch, A. C., Braun, D. C., \& Fan, Y.\yapj{2010}{723}{L190}

\bibitem[Birch et al.(2016)]{Birch16}
Birch, A. C., Schunker, H., Braun, D. C., Cameron, R., Gizon, L. L\"optien, B., \& Rempel, M.\ysci{2016}{e1600557, pp.~1--6}
{DOI:10.1126/sciadv.1600557}

\bibitem[Brandenburg(2005)]{B05}
Brandenburg, A.\yapj{2005}{625}{539}

\bibitem[Brandenburg et al.(2011)]{BKKMR11}
Brandenburg, A., Kemel, K., Kleeorin, N., Mitra, D., \& Rogachevskii, I.\yapjl{2011}{740}{L50}

\bibitem[Brandenburg et al.(2013)]{BKR13}
Brandenburg, A., Kleeorin, N., \& Rogachevskii, I.\yapjl{2013}{776}{L23}

\bibitem[Cally et al.(1994)]{CBZ94}
Cally, P. S., Bogdan, T. J., \& Zweibel, E. G.\yapj{1994}{437}{505}

\bibitem[Cally \& Bogdan(1997)]{CB97}
Cally, P. S., \& Bogdan, T. J.\yapjl{1997}{486}{L67}

\bibitem[Charbonneau(2010)]{Cha10}
Charbonneau, P.\yjour{2010}{Living Rev.\ Solar Phys.}{7}{3}

\bibitem[Cheung et al.(2010)]{CRTS10}
Cheung, M. C. M., Rempel, M., Title, A. M., \&
Sch\"ussler, M.\yapj{2010}{720}{233}

\bibitem[Chowdhury et al.(2013)]{CCG13}
Chowdhury, P., Choudhary, D. P., \& Gosain, S.\yapj{2013}{768}{188}

\bibitem[Daiffallah et al.(2011)]{DABCG11}
Daiffallah, K., Abdelatif, T., Bendib, A., Cameron, R., \& Gizon, L.\ysph{2011}{268}{309}

\bibitem[Duvall et al.(1998)]{DKM98}
Duvall, T. L., Jr., Kosovichev, A. G., \& Murawski, K.\yapj{1998}{505}{L55}

\bibitem[Fan(2001)]{Fan01}
Fan, Y.\yapj{2001}{554}{L111}

\bibitem[Felipe et al.(2012)]{FBCB12}
Felipe, T., Braun, D., Crouch, A., \& Birch, A.\yapj{2012}{757}{148}

\bibitem[Felipe et al.(2013)]{FCB13}
Felipe, T., Crouch, A., \& Birch, A.\yapj{2013}{775}{74}

\bibitem[Fernandes et al.(1992)]{FSTT92}
Fernandes, D. N., Scherrer, P. H., Tarbell, T. D., \& Title, A. M.\yapj{1992}{392}{736}

\bibitem[Getling et al.(2016)]{Get16}
Getling, A. V., Ishikawa, R., \& Buchnev, A. A.\ysph{2016}{291}{371}

\bibitem[Hanasoge et al.(2008)]{HBBG08}
Hanasoge, S. M., Birch, A. C., Bogdan, T. J., \& Gizon, L.\yapj{2008}{680}{774}

\bibitem[Ilonidis et al.(2011)]{Ilo11}
Ilonidis, S., Zhao, J., \& Kosovichev, A.\ysci{2011}{333}{993}

\bibitem[Inoue et al.(2013)]{Ino13}
Inoue, S., Hayashi, K., Shiota, D., Magara, T., \& Choe, G. S.\yapj{2013}{770}{79}

\bibitem[K\"apyl\"a et al.(2016)]{KBKKR16}
K\"apyl\"a, P. J., Brandenburg, A., Kleeorin, N., K\"apyl\"a, M. J., \& Rogachevskii, I.\yana{2016}{588}{A150}

\bibitem[Kholikov(2013)]{Kho13}
Kholikov, S.\ysph{2013}{287}{229}

\bibitem[Masada \& Sano(2016)]{MS16}
Masada, Y., \& Sano, T.\yapj{2016}{822}{L22}

\bibitem[Maurya et al.(2012)]{MVA12}
Maurya, R. A., Vemareddy, P., \& Ambastha, A.\yapj{2012}{747}{134}

\bibitem[Mitra et al.(2014)]{MBKR14}
Mitra, D., Brandenburg, A., Kleeorin, N., Rogachevskii, I.\ymn{2014}{445}{761}

\bibitem[Murawski \& Roberts(1993a)]{MR93b}
Murawski, K. and Roberts, B.\yana{1993a}{272}{601}

\bibitem[Murawski \& Roberts(1993b)]{MR93a}
Murawski, K. and Roberts, B.\yana{1993b}{272}{595}

\bibitem[Ossendrijver(2003)]{Ossen03}
Ossendrijver, M.\yanar{2003}{11}{287}

\bibitem[Schou(1999)]{Sch99}
Schou, J.\yapj{1999}{523}{L181}

\bibitem[Schrijver et al.(2011)]{Sch11}
Schrijver, C. J., Aulanier, G., Title, A. M., Pariat, E., \& Delann\'ee, C.\yapj{2011}{738}{167}

\bibitem[Singh et al.(2014)]{SBR14}
Singh, N. K., Brandenburg, A., \& Rheinhardt, M.\yapjl{2014}{795}{L8}

\bibitem[Singh et al.(2015)]{SBCR15}
Singh, N. K., Brandenburg, A., Chitre, S. M.,
\& Rheinhardt, M.\ymn{2015}{447}{3708}

\bibitem[Solov'ev \& Kirichek(2009)]{SK09}
Solov'ev, A. A., \& Kirichek, E. A.\yarep{2009}{53}{675}

\bibitem[Smirnova et al.(2013)]{Smi13}
Smirnova, V., Riehokainen, A., Solov'ev, A., Kallunki, J.,
Zhiltsov, A., \& Ryzhov, V.\yana{2013}{552}{A23}

\bibitem[Stein \& Nordlund(2012)]{SN12}
Stein, R. F., \& Nordlund, \AA.\yapjl{2012}{753}{L13}

\bibitem[Sun et al.(2015)]{SHLZ15}
Sun, X., Hoeksema, J. T., Liu, Y., \& Zhao, J.\yapj{2015}{798}{114}

\bibitem[Thomas, et al.(1982)]{TCN82}
Thomas, J. H., Cram, L. E., \& Nye, A. H.\ynat{1982}{297}{485}

\bibitem[Wachter et al.(2012)]{Wachter}
Wachter, R., Schou, J., Rabello-Soares, M. C., Miles, J. W., Duvall, T. L., \& Bush, R. I.\ysph{2012}{275}{261}

\bibitem[Warnecke et al.(2013)]{WLBKR13}
Warnecke, J., Losada, I. R., Brandenburg, A., Kleeorin, N., \& Rogachevskii, I.\yapjl{2013}{777}{L37}

\bibitem[Zhao et al.(2015)]{Zha15}
Zhao, J., Chen, R., Hartlep, T., \& Kosovichev, A. G.\yapj{2015}{809}{L15}

\end{thebibliography}
\end{document}